\documentclass[%
reprint,
 superscriptaddress,
 nofootinbib,
 amsmath,amssymb,
 aps,
]{revtex4-2}

\usepackage{graphicx}
\usepackage{dcolumn}
\usepackage{bm}

\usepackage{graphicx}
\usepackage{dcolumn}
\usepackage{bm}
\usepackage{hyperref}

\usepackage[dvipsnames]{xcolor}
\usepackage{orcidlink}
\usepackage{placeins}
\usepackage{color}
\usepackage[dvipsnames]{xcolor}
\usepackage[normalem]{ulem} 

\newcommand{\Aei}{\affiliation{Max Planck Institute for Gravitational Physics (Albert Einstein Institute), D-14467 Potsdam, Germany}}
\newcommand{\Caltech}{\affiliation{Theoretical Astrophysics 350-17, California Institute of Technology, Pasadena, CA 91125, USA}}
\newcommand{\CornellPhysics}{\affiliation{Department of Physics, Cornell University, Ithaca, NY, 14853, USA}}
\newcommand{\Cornell}{\affiliation{Cornell Center for Astrophysics and Planetary Science, Cornell University, Ithaca, New York 14853, USA}}
\newcommand{\CornellLepp}{\affiliation{Laboratory for Elementary Particle Physics, Cornell University, Ithaca, New York 14853, USA}}
\newcommand{\Fullerton}{\affiliation{Nicholas and Lee Begovich Center for Gravitational-Wave Physics and Astronomy, California State University Fullerton, Fullerton, California 92831, USA}}

\newcommand{\Perimeter}{\affiliation{Perimeter Institute for Theoretical Physics, Waterloo, ON N2L2Y5, Canada}}

\newcommand{\paperone}[0]{Paper I}
\newcommand{\paperB}[0]{Paper B}

\begin{document}


\title{
    High-accuracy drivers to simulate black hole binaries beyond general relativity \\ with the fixing-the-equations approach
    }

\author{Guillermo Lara~\orcidlink{0000-0001-9461-6292}}
\email{glara@aei.mpg.de} \Aei
\author{Harald P. Pfeiffer~\orcidlink{0000-0001-9288-519X}} \Aei

\author{Nils Deppe~\orcidlink{0000-0003-4557-4115}} \CornellLepp \CornellPhysics \Cornell
\author{Lawrence E.~Kidder~\orcidlink{0000-0001-5392-7342}} \Cornell
\author{Geoffrey Lovelace~\orcidlink{0000-0002-7084-1070}} \Fullerton
\author{Sizheng~Ma~\orcidlink{0000-0002-4645-453X}} \Perimeter
\author{Alexandra Macedo~\orcidlink{0009-0001-7671-6377}} \Fullerton
\author{Jordan Moxon~\orcidlink{0000-0001-9891-8677}} \Caltech
\author{Kyle C.~Nelli~\orcidlink{0000-0003-2426-8768}} \Caltech
\author{Mark A.~Scheel~\orcidlink{0000-0001-6656-9134}} \Caltech
\author{William Throwe~\orcidlink{0000-0001-5059-4378}} \Cornell
\author{Nils L.~Vu~\orcidlink{0000-0002-5767-3949}} \Caltech

\date{\today}

\begin{abstract}
We implement the \emph{fixing-the-equations} approach [Phys.Rev.D 96 (2017) 8, 084043] in \textsc{spectre}, an NR code using a pseudo-spectral discontinuous Galerkin scheme, to produce long and accurate NR waveforms in the well-known shift-symmetric version of scalar Gauss-Bonnet (sGB) gravity.
To achieve this, we introduce a new family of comoving driver equations that exploits the approximate symmetries of quasicircular binary systems and is designed to recover the exact (quasi-)stationary solutions of the fully-coupled theory.
We validate our single black hole (BH) solutions against analytic predictions and show that, even for binary BHs in the early inspiral, the intrinsic BH quantities are relatively insensitive to the timescales entering the driver equation.
Attention is given to the prescription of driver equations for tensors, for which we give an example of how treating tensor components as scalars can lead to undesired behaviour over long timescales, including spurious growth of the BH spins.
A more appropriate generalization to the tensor case is given for the comoving driver, which is shown to avoid these issues.
Overall, our implementation leverages state-of-the-art methods for eccentricity reduction and wave extraction with Cauchy Characteristic Evolution to simulate systems with eccentricity \(\lesssim 10^{-3}\).
We obtain waveforms with phase errors \(\lesssim \mathcal{O}(1) \, \mathrm{rad}\) over almost 40 GW-cycles, which naturally incorporate memory contributions.
\end{abstract}

\maketitle


\section{Introduction}

\begin{figure}
    \includegraphics[width=\linewidth]{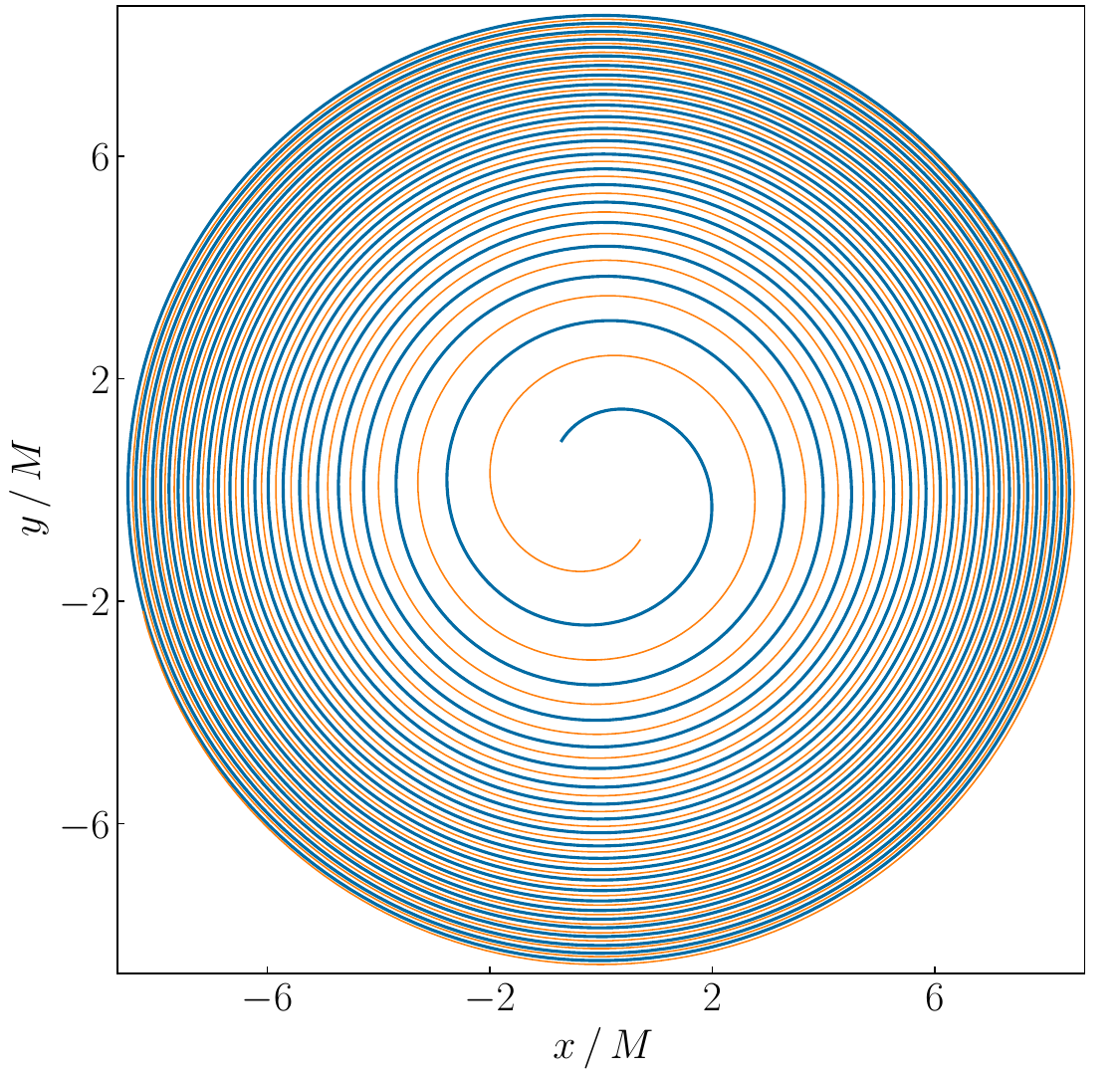}
    \caption{\emph{Binary BH inspiral in scalar Gauss-Bonnet gravity.}
    Trajectory of BHs A (blue) and B (orange) for a non-spinning equal-mass simulation with beyond-GR coupling \(\sqrt{\kappa}\ell^2/m^2 = 1/20\), where \(m\) is the initial Christodolou mass of the component BHs and \(M = 2m\).
    }
    \label{fig : inspiral illustration}
\end{figure}

Gravitational wave (GW) observations of compact objects offer a new testing ground for general relativity (GR) and numerous alternative theories of gravitation ---proposed, for instance, to explain Dark Energy or to connect gravity to quantum physics; see Refs.~\cite{Berti:2015itd, Ezquiaga:2018btd} for reviews.
Although no consensus has yet been reached on a preferred candidate theory (or class of theories) to contend against GR, developing new methods and numerical codes to compute strong-field predictions in such theories remains of general interest.
Such efforts may not only facilitate redirecting efforts should such a clear contender emerge in the future, but \emph{at present}, they can contribute to strengthening the design of tests of general relativity (TGR), and to improving forecasts for constraints on deviations from GR with next-generation GW detectors (e.g.~the Einstein Telescope~\cite{ET:2025xjr}, Cosmic Explorer~\cite{Evans:2021gyd} and the Laser Interferometer Space Antenna~\cite{LISA:2024hlh}).

In recent years, several works have shown that inspiral-merger-ringdown NR simulations are possible in various classes of theories, in particular, those where the beyond-GR corrections fundamentally change the structure of the equations of motion, affecting the properties of the related initial (boundary-) value problem (IVP) ---see e.g.~\cite{Cayuso:2017iqc, Witek:2018dmd, Okounkova:2019zep,Okounkova:2019zjf, Okounkova:2020rqw, Kovacs:2019jqj, Kovacs:2020pns, Kovacs:2020ywu, Cayuso:2020lca, East:2020hgw, Bezares:2021dma, Lara:2021piy,  Figueras:2021abd, Franchini:2022ukz, Corman:2022xqg, AresteSalo:2022hua,  Cayuso:2023xbc, Held:2023aap, Figueras:2024bba, Shum:2025lgp,  Figueras:2025wtx}.
As in the years immediately following the binary black hole (BH) breakthrough~\cite{Pretorius:2005gq, Baker:2005vv, Campanelli:2005dd}, successful evolutions have been achieved, but the resulting waveforms remain limited to few orbits and the systems explored are sparse.
With several codes now being able to perform simulations in alternative theories, the \emph{next} challenge is thus to increase the robustness, length and accuracy of the waveforms, as well as explore in a systematic way any interesting regions of parameter space (e.g.~those where ``smoking-gun'' signatures occur), at least for a few representative theories, all to allow for the most ambitious of the use cases for such waveforms.

Building full waveform inspiral-merger-ringdown models beyond GR may seem to be a daunting task.
However, for one particular theory we may be closer than one would expect to fully modeling it in this manner, though in a restricted region of parameter space.
Recently, Juli\'e, \emph{et al} have built an effective-one-body (EOB) model for scalar Gauss-Bonnet (sGB) gravity that combines results from beyond-GR extensions of Post-Newtonian and BH perturbation theory calculations ---see Ref.~\cite{Julie:2024fwy} and references therein.
While the modelled deviations from GR are restricted to nonspinning binaries, there is a realistic prospect to calibrate the model to obtain a \emph{first} full waveform model beyond GR.
Performing Bayesian inference on observed GW events using calibrated EOB models for both GR and sGB, would allow for Bayesian model selection, and for a comparison where both theories are treated on the \emph{same} footing.
Carrying out in full such an interesting exercise may well justify the effort of performing a limited simulation campaign to calibrate the model.

From the methodological point of view, the two-body problem in vacuum GR is an outstanding example of how obtaining solutions to partial differential equations with spectral methods can lead to rapid error convergence and remarkable efficiency, provided that the solutions are sufficiently \emph{smooth}. This has allowed for the construction of thousands of binary BH waveforms with the Spectral Einstein Code (\textsc{SpEC})~\cite{Mroue:2013xna, Boyle:2019kee, Scheel:2025jct}.
Even in vacuum, however, not all theories beyond GR are guaranteed to be smooth.
One example is \(k\)-essence~\cite{Armendariz-Picon:1999hyi} (a subset of Horndeski theory~\cite{Horndeski:1974wa}), in which the equation of motion for the additional scalar field is subject to shocks~\cite{Bernard:2019fjb, Bezares:2020wkn}.
Consequently, some beyond-GR theories will remain better suited for numerical evolution using finite-difference/-element or mixed methods ---see Refs.~\cite{Duez:2007bz, Duez:2008rb, Ma:2023sok, Deppe:2024ckt} for examples of the latter.

Theories that \emph{can} be simulated with spectral methods include shift-symmetric sGB gravity and dynamical Chern-Simons theory~\cite{Okounkova:2019zjf, Okounkova:2020rqw}.
The simulations of Okounkova, \emph{et al} employed a straightforward application of perturbation theory (around GR), referred to in the literature as the \emph{reduction-of-order} approach, to address issues with the well-posedness of the IVP.
The resulting waveforms were found, however, to be subject to significant accumulated secular error in the GW amplitude~\cite{Okounkova:2020rqw} ---see also Ref.~\cite{Corman:2024cdr}; see Ref.~\cite{GalvezGhersi:2021sxs} for a possible way to address secular growth using renormalization methods.

In this work, we revisit the case of sGB gravity using a different approach and implement it in our newer code \textsc{spectre}~\cite{deppe_2026_19373346} ---efforts briefly described in our companion \emph{Letter}~\cite{CompanionLetter}.
We follow the \emph{fixing-the-equations} approach~\cite{Cayuso:2017iqc} to recast the evolution system of scalar Gauss-Bonnet gravity (see also Refs.~\cite{Franchini:2022ukz, Corman:2024cdr}).
The theory is defined (in units where \(c=1\) and  \(\kappa \equiv 8 \pi G \)) by the action
\begin{align} \label{eq : action}
    S\left[g_{ab}, \Psi\right] \equiv \int d^4 x \sqrt{-g} \left[\dfrac{R}{2 \kappa} - \dfrac{1}{2} \nabla_{a} \Psi \nabla^{a} \Psi  +  \ell^{2} f(\Psi) \, \mathcal{G} \right],
\end{align}
where \(g \equiv \det (g_{ab})\) is the determinant of the metric \(g_{ab}\), \(\Psi\) is a new gravitational scalar field, and \(f(\Psi)\) is a shape function describing the coupling, with strength \(\ell\), to the Gauss-Bonnet scalar
\begin{align}
    \mathcal{G} &\equiv R_{abcd} R^{abcd} - 4 R_{ab} R^{ab} + R^{2} ,
\end{align}
which is in turn defined in terms of contractions of the Riemann tensor \(R_{abcd}\), Ricci tensor \(R_{ab} \equiv g^{cd} R_{cadb}\), and Ricci scalar \(R \equiv g^{cd} R_{cd}\).
We further simplify to the case of the shift-symmetric version of the theory, \(f(\Psi) = \Psi\), in which all BHs have non-trivial secondary \emph{hair}~\cite{Sotiriou:2013qea, Sotiriou:2014pfa}. 
(Here, we will only consider the simplest type of \emph{scalarized} BHs; see Ref.~\cite{Doneva:2022ewd} for a review of other scalarization mechanisms for which the present work can be straightforwardly extended.)
The coupling scale for this case is constrained by parameterized LVK tests which yield \(\ell/2 = \sqrt{\alpha_{\mathrm{GB}}} \lesssim 0.3 \, \mathrm{km}\)~\cite{Sanger:2024axs}.

In order to address issues with the well-posedness of the initial value problem, the \emph{fixing-the-equations} approach~\cite{Cayuso:2017iqc} prescribes a way to promote certain beyond-GR terms in the equations of motion to become auxiliary dynamical variables, all with the objective of bringing the system to a form in which known methods can be applied. 
Such variables are subject to \emph{ad hoc} evolution equations that \emph{drive} them towards target values, given by the original expressions for the beyond-GR corrections, on prescribed timescales.
In the context of BH simulations, the method essentially prescribes a way to ``filter'' the irrelevant short wavelength modes (shorter than the BH or GW scales) that may render the evolution unstable.
Another intuitive way to interpret the method, when we regard sGB as an effective field theory (EFT), is that the \emph{fixing-the-equations} method allows us to implement a version of the theory where the truncation (here at 4th order in a derivative expansion) is ``softened'' by the auxiliary variables ---see e.g.~Refs.~\cite{Allwright:2018rut,Lara:2024rwa} for an illustrative example in which the ``fixed'' theory essentially reintroduces an ultraviolet degree of freedom that had been integrated out to obtain the truncated theory; see also Ref.~\cite{Besharat:2026rba} for an example where energy is dissipated into an additional spacetime dimension.
The method, however, does \emph{not} specify the particular form that the driver equations must take, except for the minimal requirements that the auxiliary variables are driven to the appropriate target, and that the driver equation does not introduce further challenges to obtain a well-behaved evolution system.

Concretely, here we continue the work of Ref.~\cite{Lara:2024rwa} (hereinafter referred to as \paperone), in which we proposed a scalar driver equation that effectively recovered the exact stationary scalar field configurations of isolated BHs in the test-field approximation.
An important contribution of \paperone~was the realization that comoving derivatives can be used to prescribe driver equations that are efficient both for static BHs and for BHs with linear momentum. 
In this paper, we make use of the approximate helical symmetry of the system (and its associated approximate Killing vector) to compute the comoving derivative terms and make the driver equation efficient for BHs in quasicircular motion.
We have also generalized such comoving drivers to the case of tensor auxiliary variables; in general, applying the \emph{fixing-the-equations} method for a generic theory beyond GR may require multiple scalar, vector and tensor auxiliary variables.

Remarkably, we have found a pathology in certain forms of the tensor driver equation.
In Sec.~\ref{sec : Results binary bbh CD1}, we illustrate how a straightforward application of the scalar driver equation to the individual tensor components of the tensor auxiliary variables in the binary case (c.f.~Ref.~\cite{Corman:2022xqg}), while seemingly well-behaved at first, can lead to spurious growth of the BH spins over long evolution times.
We correct this behaviour by giving a vector/tensor generalization that is \emph{free} of this problem and which leads to improved performance.
The \emph{key} insight is to write the comoving derivatives in the tensor equation in terms of Lie derivatives, with the objective of properly Lie-dragging the auxiliary tensor variable along the orbital motion.

In summary, the \emph{fixing-the-equations} prescription given here minimizes two possible sources of systematic error: \emph{i)} the error on the intrinsic description of the component BHs (addressed by the comoving derivatives), and \emph{ii)} the error due to orbital variation of the auxiliary variables in the inertial frame (addressed by the Lie derivatives).

On the numerical implementation side, we build on recent progress~\cite{Lovelace:2024wra} in the development of \textsc{spectre} to perform long inspiral simulations for equal mass systems.
Eccentricity-reduction is carried out using methods developed for GR, and we find that they are generally effective in reducing the eccentricity to \(\lesssim 10^{-3}\) for scalarized binaries.
Note that these are not guaranteed \emph{a priori} to work as the development of these methods did not consider any additional radiation channels.
The gravitational and scalar waves are extracted at future-null infinity using state-of-the-art Cauchy Characteristic Evolution (CCE)~\cite{Moxon:2021gbv} (and its generalization for scalar-tensor theories~\cite{Ma:2024bed}), which results in waveforms that include memory.
We quantify the phase accuracy of our waveforms both in resolution and in their dependence with the auxiliary timescales introduced by the \emph{fixing-the-equations} approach.
For a benchmark binary system with small coupling, the phase error due to the choice of auxiliary timescales is lower than the numerical resolution error, and of order \(\mathcal{O}(1 \, \mathrm{rad})\).
Taken together, these results represent substantial advances in our efforts to obtain waveforms that can be used for the numerous applications regarding TGR.

This paper is organized as follows.
In Sec.~\ref{sec :  Theory}, we recap the basic equations and outline the formulation of the evolution system using the \emph{fixing-the-equations} approach.
Our numerical implementation as well as the procedure for eccentricity reduction and wave extraction is described in Sec.~\ref{sec : Methodology}.
Tests for single BHs and validation with theoretical predictions are shown in Sec.~\ref{sec : Results Single Black Holes}, where we highlight how the \emph{comoving} drivers recover the stationary solutions of the theory.
We then proceed to report the main results of this paper for the full binary BH problem.
In Sec.~\ref{sec : Results binary bbh CD1} we discuss scalar drivers and the spin growth issue, which is then resolved in Sec.~\ref{sec : Results binary bbh CD2} by employing the CD2 generalization of the tensor driver.
We discuss the obtained waveforms in Sec.~\ref{sec : bbh results extracted waves}.
Further discussion, including limitations and relation to other work, is presented in Sec.~\ref{sec : Discussion}, with concluding remarks in Sec.~\ref{sec : Conclusions}.
More technical details are given in the appendices:
a comparison between our comoving drivers and the advection driver of Ref.~\cite{Cayuso:2023xbc} is shown in App.~\ref{app : appendix comparison with and advection driver};
we find no evidence of secular growth in the amplitude of the waveform in App.~\ref{app : waveform length dependence};
and details on the evolution system are given in Apps.~\ref{app : details first order system} and \ref{app : source term expressions}.
Throughout this paper we use \((-+++)\) signature. We denote spacetime indices with early alphabetic letters \(\{a, b, c, \dots\}\), and spatial indices with middle alphabetic letters \(\{i, j, k, \dots\}\).


\section{Theory \label{sec :  Theory}}

Variation of the action [Eq.~\eqref{eq : action}] yields the equations of motion,
\begin{align} \label{eq : equations of motion}
    \Box \Psi &= \mathcal{S}\left[\partial^{\leq 2} g \right] , \nonumber \\
    R_{ab} &= \mathcal{S}_{ab}\left[\partial^{\leq 2} g, \partial^{\leq 2} \Psi \right] ,
\end{align}
where the source terms depend on up to second-order derivatives of the metric \(g_{ab}\) and the scalar field \(\Psi\), and are given by
\begin{align} \label{eq : source terms abstract}
    \mathcal{S} &\equiv - \ell^2 f'(\Psi) \, \mathcal{G} , \nonumber \\
    \kappa^{-1}\mathcal{S}_{ab} &\equiv T^\mathrm{(TR)}_{ab} + \ell^2 H^\mathrm{(TR)}_{ab} .
\end{align}
Here we have defined
\begin{align} \label{eq : THP definitions}
    T_{ab} &\equiv \nabla_{a}\Psi \nabla_{b} \Psi - \dfrac{1}{2} \nabla_{c} \Psi \nabla^{c} \Psi g_{ab} , \nonumber \\
    H_{ab} &\equiv -8 P_{acbd} \nabla^{c} \nabla^{d} f(\Psi) ,\nonumber \\
    P_{abcd} &\equiv R_{abcd} - 2 g_{a\left[c\right.} R_{\left.d\right] b} + 2 g_{b \left[c\right.} R_{\left.d\right] a} + g_{a\left[c\right.}g_{\left.d\right] b} R . 
\end{align}
We use the (TR) label on second-rank tensors to refer to the tensor resulting from applying the trace-reverse operation, i.e.~for \(w_{ab}\), the trace-reversed tensor is \(w^{\mathrm{(TR)}}_{ab} \equiv w_{ab} - (1/2) g_{cd}w^{cd}g_{ab}\).

In this paper we will focus on the shift-symmetric version of the theory, given by
\begin{align} \label{eq : sGB shape function}
    f(\Psi) \equiv \Psi.
\end{align}
This is the only type of coupling that leads to BHs with nontrivial \emph{scalar hair} and maintains shift-symmetry~\cite{Sotiriou:2013qea, Sotiriou:2014pfa, Capuano:2023yyh}.


\subsection{Fixing the equations \label{sec : Theory Formulation}}

We follow the \emph{fixing-the-equations} approach~\cite{Cayuso:2017iqc} to rewrite the  system~\eqref{eq : equations of motion} in a way that is suitable for numerical evolution.
Here we will briefly review how we implemented this approach in the test-field approximation (\paperone), and provide the necessary extensions to account for the full system of equations~\eqref{eq : equations of motion}.

As in \paperone, we replace the source terms \(\boldsymbol{\mathcal{S}} \equiv (\mathcal{S}, \mathcal{S}_{ab})\) by auxiliary variables \(\boldsymbol{\Sigma} \equiv (\Sigma, \Sigma_{ab})\) in the right-hand-side (RHS) of Eq.~\eqref{eq : equations of motion}, i.e.,
\begin{align} \label{eq : fixed equations of motion}
    \Box \Psi &= \Sigma , \nonumber \\
    R_{ab} &= \Sigma_{ab} .
\end{align}
The variables $\boldsymbol\Sigma$ are evolved according to \emph{ad hoc}, second-order driver equations, represented abstractly by
\begin{align} \label{eq : driver equation abstract}
    \boldsymbol{\mathcal{F}}[\partial^{\leq 2} \boldsymbol{\Sigma}, \boldsymbol{\mathcal{S}}; \boldsymbol{\lambda}] = 0 .
\end{align}
Equation~\eqref{eq : driver equation abstract} is prescribed such that \(\boldsymbol{\Sigma}\) closely \emph{tracks} \(\boldsymbol{\mathcal{S}}\) in a way that depends on specifiable auxiliary timescales \(\boldsymbol{\lambda}\).

As a simple example consider a driver for a single degree of freedom, \(y(t)\).
One of the simplest forms of Eq.~\eqref{eq : driver equation abstract}  is \(\tau \partial_t y(t) = - \left[y(t) - s(t)\right]\), where \(\boldsymbol{\lambda} = \{\tau\}\) contains a single positive timescale \(\tau\).
For \(s(t) =  \mathrm{const.} \), one can see that \(y(t) = y(t_0) e^{-(t-t_0)/\tau} + s_0 (1-e^{-(t-t_0)/\tau}) \) approaches \( s_0 \) exponentially as \(t \to \infty\).
For time-varying \(s(t)\), we have [c.f.~Eq.~(9) of Ref.~\cite{Lindblom:2007xw}]
\begin{align} \label{eq : simple example full solution}
    y(t) = y(t_0) e^{-(t-t_0)/\tau} + \dfrac{1}{\tau} \int_{t_0}^{t} dt' \, e^{-(t-t')/\tau} s(t'),
\end{align}
where the second term is a smoothed-out version of \(s(t)\) given by an exponentially weighed average.
More generally, the role of driver equation is to make \(\boldsymbol{\Sigma}\) an approximation of the source terms \(\boldsymbol{\mathcal{S}}\) where high-frequency modes have been filtered out.
For the example above, the low-pass filtering is carried out by the second term in Eq.~\eqref{eq : simple example full solution}.

In the context of BH evolution, the auxiliary timescales \(\boldsymbol{\lambda}\) should be chosen to be small enough to properly account for all relevant scales in the system (roughly \(\lesssim m_2\), where \(m_2\) is the mass of the smallest BH), while discarding the high-frequency modes that are \emph{not} relevant for the problem, but which could nevertheless potentially lead to instabilities in the solutions.
The expectation is that by keeping the relevant frequency modes, the predictions made with this approach are physically meaningful, and relatively insensitive to the particular choice of driver equation [c.f.~\eqref{eq : driver equation abstract}], provided that \(\boldsymbol{\lambda}\) are chosen appropriately.
See Ref.~\cite{Cayuso:2017iqc} for a more in-depth discussion and for a description of potential exceptions to these expectations.
In practice, we perform numerical evolutions with different values for the driver timescales in order to assess the consistency of the resulting solutions, similarly to the way we study the dependence on numerical resolution.

Recalling that the principal part consists of the terms with the highest derivative order, we notice that in writing the equations of motion in the form~\eqref{eq : fixed equations of motion}, we have split the principal part into two types of terms.
The first type are the terms on the left-hand-side (LHS) of Eq.~\eqref{eq : fixed equations of motion}, which are reminiscent of equations of motion of GR minimally-coupled to a scalar and can be handled with the standard formulations and gauges developed for GR.
The second type, on the RHS, are the beyond-GR corrections to the principal part of order \(\mathcal{O}(\ell^2)\), which could in principle alter the character of the equation if they become sufficiently large (from hyperbolic to elliptic) ---see e.g.~Ref.~\cite{Ripley:2019hxt}.
This split is justified by adopting the viewpoint that the theory described by action~\eqref{eq : action} can be regarded as an effective field theory, a special case of the more general scalar-tensor EFT of Ref.~\cite{Weinberg:2008hq}.
From this perspective, terms of order \(\mathcal{O}(\ell^2)\) are expected to be  subleading for the theory to remain consistent.
In Sec.~\ref{sec : closures}, we again make use of this assumption to give a prescription on how to compute the source terms \(\boldsymbol{\mathcal{S}}\).


\subsection{Comoving drivers \label{sec : comoving driver}}

Several choices for the driver equation~\eqref{eq : driver equation abstract} have been considered in the literature, including exponential equations (e.g.~Ref.~\cite{Allwright:2018rut}), wavelike equations (e.g.~Ref.~\cite{Franchini:2022ukz}), and advective equations (e.g.~Ref.~\cite{Cayuso:2023xbc}) ---variants of these equations were also used before as \emph{gauge} drivers (see e.g.~Ref.~\cite{Lindblom:2007xw}) or in the context of dissipative hydrodynamics~\cite{Cayuso:2017iqc}.
Some of these equations, however, struggle to recover the exact stationary solutions of the theory, recovering them only in the limit in which the auxiliary timescales \(\boldsymbol{\lambda}\) approach zero.
This is the case, for instance, for wavelike equations in which spatial derivative terms do \emph{not} vanish (or cancel) when all fields are stationary ---we will come back to this point in Sec.~\ref{sec : Discussion}.

With this issue in mind, and with the benefit of hindsight, we specify Eq.~\eqref{eq : driver equation abstract} as
\begin{align} \label{eq : comoving driver}
    (\sigma \mathcal{D}^2_{t} + \tau \mathcal{D}_{t} ) \boldsymbol{\Sigma} = - \left[\boldsymbol{\Sigma} - \boldsymbol{\mathcal{S}}\right],
\end{align}
where \(\boldsymbol{\lambda} = (\sigma, \tau)\) are positive dimensionful parameters, with units of \([\mathrm{time}]^2\) and \([\mathrm{time}]\), respectively.
We define a \emph{comoving} derivative operator 
\begin{align}
    \mathcal{D}_t \equiv \mathcal{L}_{\xi},
\end{align}
in terms of the Lie derivative with respect to a vector \(\xi^{a}\).
(In fluid dynamics, these types of operators are sometimes called \emph{material} derivatives and \(\xi^{a}\) corresponds to the flow velocity.)
To fully specify \(\mathcal{D}_t\), we can take advantage of any (approximate) symmetries of the spacetime, and define \(\xi^{a}\) to be the associated (approximate) timelike Killing vector.
If such (approximate) symmetries exist, we expect that physical quantities and fields \(\mathcal{Q}\) will at worst vary slowly with respect to this operator, i.e.~\(\mathcal{D}_t \mathcal{Q} \approx 0\).
For example, in the case of an isolated BH with constant velocity \(v^{a}_\mathrm{BH}\), the Killing vector is the boosted vector \({\chi^{a}=\Lambda(v_\mathrm{BH})^{a}}_{b}(\partial_t)^{b}\).
For non-spinning quasicircular BH binaries, one can identify a helical vector that describes the approximate symmetry of the orbital motion.

In the following, we do not attempt to construct the closest approximation to the Killing vectors for dynamical spacetimes.
Instead, we restrict to vectors of the form \(\chi^{a} = (1, v^{i})\) to define \(\mathcal{D}_t\).
Here,
\begin{align} \label{eq : mesh velocity}
    v^{i} \equiv \dfrac{\partial x^{i}}{\partial \hat{t}}
\end{align}
is the \emph{frame velocity} relating the inertial coordinates \(\{t = \hat{t}, x^{i}(\hat{t}, \hat{x}^{i})\}\) and a set of \emph{comoving} coordinates \(\{\hat{t}, \hat{x}^{i}\}\).
(Conveniently, our NR code, \textsc{spectre}, already constructs such mappings by following the moving-frames approach~\cite{Scheel:2006gg}.)

For a quasicircular BH binary, the comoving coordinates we will use are essentially those of a co-rotating frame (and since the mapping between these coordinates is foliation-preserving, they do not take into account possible boost transformations).
Explicitly, \(\mathcal{D}_t\) acts on the components of \(\boldsymbol{\Sigma}\) as
\begin{align}\label{eq : scalar comoving operator}
    \mathcal{D}_t \Sigma &= \left(\partial_t + v^{i} \partial_{i}\right) \Sigma ,\\
    \label{eq : tensor comoving operator}
    \mathcal{D}_t \Sigma_{ab} &=  \left(\partial_t + v^{i} \partial_{i}\right) \Sigma_{ab} + 2\delta^{0}_{(a}\Sigma_{b)i}\partial_t v^{i} \nonumber \\
    &\quad\qquad\qquad\qquad\qquad\qquad + 2\delta^{i}_{(a}\Sigma_{b)j}\partial_{i}v^{j}.
\end{align}

In \paperone, we proposed a first version for the scalar part of Eq.~\eqref{eq : comoving driver}, which originally also included advective terms of the form \((\partial_t - \beta^{i}\partial_{i})\), where \(\beta^{i}\) is the shift vector.
Here, after further experimentation, we have found that expressing the driver fully in terms of \(\mathcal{D}_t\) leads to better tracking properties for the two-body problem.
Noticing that Eq.~\eqref{eq : comoving driver} is reminiscent of the damped harmonic oscillator in the comoving frame, we introduce the dimensionless damping parameter \(\zeta_d  \equiv \tau / (2 \sqrt{\sigma})\) to label our parameter choices in terms of \(\{\sigma, \zeta_d\}\) instead of \(\{\sigma, \tau\}\).
Unless stated otherwise, we will only consider the critically-damped case, \(\zeta_d = 1\).


\subsubsection{Simplified comoving drivers}

Given how \(\mathcal{D}_t\) acts on tensors, Eq.~\eqref{eq : comoving driver} may seem rather complicated.
We will see below, however, that most of this complexity is justified.
The main difficulty in the practical implementation of Eq.~\eqref{eq : comoving driver} in our code is computing \(\partial_t v^{i}\), which requires computing the Hessian of a large composition of maps that rotate, scale, translate and distort the numerical grid.
Computing \(\partial_{j}v^{i}\) is easier, as we can obtain it by evaluating numerical derivatives; see Sec.~\ref{sec : first order system and numerical implementation} for more details.

To appreciate the need for the complexity in Eq.~\eqref{eq : comoving driver}, and to ease some of the difficulties in its implementation, we will consider two \emph{simplified} versions of Eq.~\eqref{eq : comoving driver}.
In the first version of the comoving driver (named CD1), we treat the tensor variables as if they were scalars,
\begin{align} \label{eq : comoving driver simplified}
\sigma (\partial_t + v^{i} \partial_{i})^2 \boldsymbol{\Sigma} + \tau (\partial_t + v^{i} \partial_{i}) \boldsymbol{\Sigma} = - \left[\boldsymbol{\Sigma} - \boldsymbol{\mathcal{S}}\right].
\end{align}
In the second version (named CD2), we keep \(\partial_j v^{i}\) terms in Eq.~\eqref{eq : comoving driver}, but set \(\partial_t v^{i}\) to zero.
Namely,
\begin{align} \label{eq : comoving driver simplified second version}
    \sigma (\partial_t + \mathcal{L}_{v})^2 \boldsymbol{\Sigma} + \tau (\partial_t + \mathcal{L}_{v}) \boldsymbol{\Sigma} = - \left[\boldsymbol{\Sigma} - \boldsymbol{\mathcal{S}}\right],
\end{align}
where \(\mathcal{L}_{v}\) is the Lie derivative with respect to the spatial vector \(v^{i}\).
Explicitly, \((\partial_t + \mathcal{L}_{v})\) acts on \(\boldsymbol{\Sigma}\) as
\begin{align}
&    (\partial_t + \mathcal{L}_{v}) \Sigma&&\hspace*{-1.6em}= \left(\partial_t + v^{i} \partial_{i}\right) \Sigma ,\nonumber \\
&    (\partial_t + \mathcal{L}_{v}) \Sigma_{00} &&\hspace*{-1.6em}= \left(\partial_t + v^{i} \partial_{i}\right) \Sigma_{00} ,\nonumber \\
&    (\partial_t + \mathcal{L}_{v}) \Sigma_{0i} &&\hspace*{-1.6em}=  \left(\partial_t + v^{i} \partial_{i}\right) \Sigma_{0i} + \Sigma_{0j}\partial_{i}v^{j}, \nonumber \\
&    (\partial_t + \mathcal{L}_{v}) \Sigma_{ij} &&\hspace*{-1.6em}=  \left(\partial_t + v^{i} \partial_{i}\right) \Sigma_{ab} + 2\Sigma_{k(i}\partial_{j)}v^{k},
\end{align}
where \(\{\Sigma, \Sigma_{00}\}\) are treated as scalars, \(\Sigma_{0i}\) as a spatial vector, and \(\Sigma_{ij}\) as a spatial tensor.
The approximation behind CD2 is motivated by the observation that in the early inspiral of a quasicircular BH binary, the orbital frequency \(\Omega_\mathrm{orb}\) varies on the radiation-reaction (RR) timescale, which in the early inspiral is much slower than the orbital timescale.
Since \(v^{i}(t, x) \sim (\boldsymbol{\Omega}_\mathrm{orb} \times \boldsymbol{x})^{i}\), the frame-velocity field also varies on the same RR timescale.
Below, we will specify \(\{\sqrt{\sigma}, \tau\} \) to be parameters with scales comparable to BHs horizon size. 
Thus, \(v^{i}\) will vary slowly with respect to the timescales on which the driver equations act, justifying setting its time derivative to zero.
Another important property of CD2, compared to the first simplification [Eq.~\eqref{eq : comoving driver simplified}], is that it takes into account the variation in the tensor components of \(\Sigma_{ab}\) introduced by orbital motion, which is correlated to keeping the BH spins under control for inspiral evolutions.


\subsubsection{Comoving drivers for puncture methods}

It may not be necessary to use a moving-mesh approach to employ the comoving drivers  described above.
One may use the puncture trajectories to construct approximations for the frame-velocity fields needed to evaluate the terms in our comoving driver proposals [Eq.~\eqref{eq : comoving driver simplified} and Eq.~\eqref{eq : comoving driver simplified second version}].
The puncture trajectories can be obtained from
\(\dot{\boldsymbol{r}}_{A} = - \boldsymbol{\beta}(\boldsymbol{r}_{A})\) and \(\dot{\boldsymbol{r}}_{B} = - \boldsymbol{\beta}(\boldsymbol{r}_{B})\), for BHs \(A\) and \(B\), where \(\boldsymbol{\beta}\) is the shift vector~\cite{Campanelli:2005dd}.
Then, we can construct the velocity field (and its derivatives) as
\begin{align}
    v^{i}(t, x) &= {\epsilon^{i}}_{jk} \Omega_{\mathrm{orb}}^{j} (x^{k}-r^{k}_{\mathrm{CM}}) + \dot{a}(x^{i} - r^{i}_{\mathrm{CM}}) , \nonumber\\
    \partial_{k}v^{i}(t, x) &= {\epsilon^{i}}_{jk}\Omega_{\mathrm{orb}}^{j} + \dot{a} \delta^{i}_{k} , \nonumber \\
    \partial_t v^{i}(t, x) &= {\epsilon^{i}}_{jk}\dot{\Omega}_{\mathrm{orb}}^{j}(x^{k}-r^{k}_{\mathrm{CM}}) + \ddot{a} (x^{i}-r^{k}_{\mathrm{CM}}) ,
\end{align}
where the separation, orbital velocity, radial expansion and center of mass (CM) are given by
\(\boldsymbol{r}_{AB} \equiv \boldsymbol{r}_{B} - \boldsymbol{r}_{A}\),
\(\boldsymbol{\Omega}_{\mathrm{orb}} \equiv (\boldsymbol{r}_{AB} \times \dot{\boldsymbol{r}}_{AB})/r^2_{AB}\),
\( \dot{a} \equiv ({\boldsymbol{r}}_{AB}\cdot \dot{\boldsymbol{r}}_{AB})/r^2_{AB}\),
and \(\boldsymbol{r}_{\mathrm{CM}} \equiv (m_A \boldsymbol{r}_A + m_B \boldsymbol{r}_{B}) / M \),
respectively, and \({\epsilon^{i}}_{jk}\) is the Levi-Civita symbol.
We will however, not explore this possibility here.


\subsection{Generalized harmonic gauge and constraint propagation \label{sec : gh and constraint propagation}}

By writing the equations of motion in the form~\eqref{eq : fixed equations of motion}, we have made it possible to apply the standard methods of NR to evolve the system.
For the metric sector, we will use a generalized harmonic formulation~\cite{Friedrich:1985afv,Lindblom:2005qh}.
Below, we briefly recap this formulation and describe how constraint propagation differs from the GR case when using the \emph{fixing-the-equations} approach, and notably, how non-zero constraint-violations may arise in this approach.

In the generalized harmonic gauge, the coordinates satisfy the harmonic condition \(\Box x^{c} = -\Gamma^{c} = H^{c}\), with gauge source \(H^{c}\)
and \(\Gamma^{c} \equiv g^{ab}\Gamma^{c}_{ab}\) defined as a contraction of the 4-dimensional Christoffel symbol \(\Gamma^{c}_{ab}\).
To impose the gauge, we rewrite the metric equations as
\begin{align} \label{eq : relaxed einstein equations}
    R_{ab} + 2 \nabla_{(a} \mathcal{C}_{b)} = \Sigma_{ab} ,
\end{align}
where
\begin{align} \label{eq : gauge constraint}
    \mathcal{C}^{c} \equiv  H^{c} + \Gamma^{c},
\end{align}
is a gauge constraint.
The original metric equations are recovered when \(\mathcal{C}^{a} \equiv 0\), i.e.~when the gauge condition is satisfied.
In order for the principal part on the LHS of Eq.~\eqref{eq : relaxed einstein equations} to be manifestly hyperbolic, i.e.~\(g^{cd} \partial_{c} \partial_{d} g_{ab} \simeq 0 \), 
the gauge source \(H_{c}\) is allowed to depend on the evolution fields, but not on their derivatives ---e.g.~the damped harmonic gauge~\cite{Choptuik:2009ww, Szilagyi:2009qz, Deppe:2018uye}, which we will use below.
(Here, we have used the notation \(\simeq\) to ignore terms that are not in the principal part.)
For the scalar equation [Eq.~\eqref{eq : fixed equations of motion}] and the LHS of the driver [Eq.~\eqref{eq : comoving driver}], the principal parts of the equations are also already manifestly hyperbolic as they are either wave or advective equations.

From the Bianchi identities,
\begin{align}\label{eq : gauge constraint propagation}
    \nabla^{a} \nabla_{a} \mathcal{C}_{a} + \mathcal{C}^{a} \nabla_{(a}\mathcal{C}_{b)} = - \kappa \Sigma_{ab}  \mathcal{C}^{a} 
    - 2 \kappa \nabla^{a} \Sigma^\mathrm{(TR)}_{ab}.
\end{align}
In GR, diffeomorphism-invariance ensures that the stress-energy tensor is conserved, i.e.~\(\nabla^{a} \Sigma^\mathrm{(TR)}_{ab} \to \nabla^{a} T_{ab} = 0\). Thus Eq.~\eqref{eq : gauge constraint propagation} ensures that, once we impose
\(\mathcal{C}^{c} \rvert_{t=t_0} = \partial_t \mathcal{C}^{c} \rvert_{t=t_0} = 0\) at initial-time,
\(\mathcal{C}^{a} \equiv 0\) is preserved.

By contrast, in the \emph{fixing-the-equations} approach, Eq.~\eqref{eq : relaxed einstein equations} does not arise from an action, and therefore, diffeomorphism-invariance does not ensure that the (trace-reversed) driver tensor \(\Sigma_{ab}\), is conserved.
Instead, we expect that the driver tensor is only \emph{approximately} conserved
\begin{align}\label{eq : approx conserved sigma}
    \nabla^{a} \Sigma^\mathrm{(TR)}_{ab} \approx 0,
\end{align}
and that small constraint-violations arise from  Eq.~\eqref{eq : gauge constraint propagation}.
The timescales in the driver equation and the concrete choice of driver equation [e.g.~Eqs.~\eqref{eq : comoving driver simplified}, \eqref{eq : comoving driver simplified second version} or \eqref{eq : wavelike driver}] will control how good the approximation is.
By decreasing the timescales we expect that the approximation improves and the constraint-violations are reduced as well ---see Refs.~\cite{Lara:2021piy,Corman:2024cdr} for similar discussions.

\emph{If} the driver timescales \emph{cannot} be arbitrarily reduced (either for practical issues, such as the equations becoming stiff, or due to fundamental reasons), this would constitute a fundamental \emph{limitation} in the accuracy of the method.
In such a situation, increasing the numerical resolution will eventually resolve such \emph{non-zero} gauge constraint violations.
For instance, Ref.~\cite{Cayuso:2017iqc} explored toy problems where the \emph{fixing-the-equations} timescales were limited by the coupling scale of the theory.
As we will see below, we find empirically that sGB BH evolutions may behave similarly (see also Ref.~\cite{Franchini:2022ukz}), however, with constraint violations small enough still allow us to produce informative evolutions;
whether there is a fundamental reason for this has not been clearly established.


\subsection{Closing the system \label{sec : closures}}

As we have shown above, the source terms \(\boldsymbol{\mathcal{S}}\) in the RHS of the original system~\eqref{eq : equations of motion}, which appear in the RHS of the driver equation~\eqref{eq : comoving driver}, depend on up to second-order derivatives of both of the dynamical fields of the original theory, \(g_{ab}\) and \(\Psi\).
To close the system, we therefore need a prescription for how to compute them.
Specifically, we need to prescribe how to compute the terms containing second-order \emph{time} derivatives; second-order spatial or mixed derivatives are not an issue as they can be computed with discretized operators.

By treating the theory as an EFT, we simply replace \(\{\partial^{2}_t \Psi, R_{ab}\}\) in \(\boldsymbol{\mathcal{S}}\) with Eq.~\eqref{eq : fixed equations of motion}, as we would do in a perturbative approach ---see App.~\ref{app : source term expressions} for the explicit expressions.
There is, however, a second possibility:
one could also replace the terms containing \(\{\partial^{2}_t \Psi, {R_{ab}}\}\) iteratively, starting with the equations of motion at order \(\mathcal{O}(\ell^0)\).
We have also explored the latter in some simple cases with similar results, and have thus chosen to focus here on the first type of closure.

\section{Methodology \label{sec :  Methodology}}


\subsection{First-order system and numerical implementation \label{sec :  first order system and numerical implementation}}

We recast Eqs.~\eqref{eq : fixed equations of motion} and Eq.~\eqref{eq : comoving driver simplified} as a first-order system of the form
\begin{align} \label{eq : first order system}
    \partial_t \boldsymbol{u} + \mathbb{A}^{i} \left(\boldsymbol{u}\right) \partial_i \boldsymbol{u} = \boldsymbol{S}\left(\boldsymbol{u}\right).
\end{align}
Here, \(\mathbb{A}^{i}\left(\boldsymbol{u}\right)\) is the principal-part matrix, \(\boldsymbol{{S}}\left(\boldsymbol{u}\right)\) is the collection of all source terms, and \(\boldsymbol{u}\) is a collection of first-order variables defined in terms of \(\{g_{ab}, \Psi, \Sigma, \Sigma_{ab}\}\) and their first derivatives ---see Apps.~\ref{app : details first order system} and \ref{app : source term expressions} for the details of this system.
We write the equations of motion for the metric using the first-order generalized harmonic system of Ref.~\cite{Lindblom:2005qh} and specify the damped harmonic gauge condition~\cite{Choptuik:2009ww, Szilagyi:2009qz, Deppe:2018uye}.
For the scalar and driver variables, we use similar first-order formulations of the evolution equations.
We include all the source terms appearing in the full equations of motion~\eqref{eq : fixed equations of motion}.

The system~\eqref{eq : first order system} is implemented in \textsc{spectre}~\cite{deppe_2026_19373346}. 
The spatial discretization is done using a discontinuous-Galerkin scheme~\cite{Teukolsky:2015ega} and the BH interior is excised from the numerical domain.
Integration in time is done following the method of lines using a 4th-order Adams-Moulton adaptive local time-stepping integrator.
As in Ref.~\cite{Scheel:2006gg}, we also use a moving-mesh approach.
The inertial coordinates \(\{t = \hat{t}, x^{i}(\hat{t}, \xi^{\hat{i}})\}\)
are related through a composition of several time-dependent mappings to the logical coordinates \(\{\hat{t}, \xi^{\hat{i}}\}\) used in the discontinuous-Galerkin scheme, and with respect to which the excision spheres are at fixed locations.
The mesh velocity \(v^{i}\) is obtained by using these maps in Eq.~\eqref{eq : mesh velocity}.
Control systems dynamically adjust these coordinate mappings
such that the computational grid follows the motion and deformation of the BHs~\cite{Hemberger:2012jz, Nelli:2025vfc}.
We evaluate the LHS of the comoving driver [Eq.~\eqref{eq : comoving driver simplified} or Eq.~\eqref{eq : comoving driver simplified second version}] using Eq.~\eqref{eq : mesh velocity} and numerical spatial derivatives of it.

The constraint damping parameters \(\gamma_{0, 1, 2}\) for the metric are given a spatial dependence described by a superposition of Gaussian functions that move with the BHs, and described in detail in Ref.~\cite{Lovelace:2024wra}.
For the scalar system's constraint damping parameter \(\gamma^{(\Psi)}_{2}\), we adopt a spatial dependence analogous to that of \(\gamma_{2}\), and we set \(\gamma^{(\Psi)}_1 = 0\).
The driver parameters \(\{\sigma, \tau\}\) in Eq.~\eqref{eq : comoving driver simplified} are given a single-Gaussian spatial dependence, described in \paperone~[Eq.~(27)].
Hereinafter, values quoted for these parameters correspond to their values at the center of the Gaussian profile. 
We refer the reader to Ref.~\cite{Lovelace:2024wra} for a more detailed discussion of the methods and spatial domain decomposition for binary black holes in GR,
most of which have been straightforwardly extended to account for the additional fields considered here.


\subsection{Initial data, ramp up and eccentricity reduction \label{sec : initial data ramp up and eccentricity reduction}}

In the single BH case, we simply evaluate the Kerr solution in Cartesian Kerr-Schild coordinates (see e.g.~\paperone).
For binary BH data we solve the extended conformal thin-sandwich (XCTS) equations using the elliptic solver within \textsc{spectre}~\cite{Vu:2021coj, Vu:2024cgf, Mendes:2025gov}.
In this way, we obtain quasistationary BH initial data in GR ---quasistationary data for BH binaries in the fully-coupled sGB theory have not been constructed yet; see Refs.~\cite{Kovacs:2021lgk, Brady:2023dgu, Nee:2024bur} for recent efforts in this direction.

We track the individual BH properties by computing quasi-local quantities using the standard formulas from GR.
These include horizon area \(A\),
angular momentum \(\boldsymbol{S}\) [Eqs.~(3)-(4) of Ref.~\cite{Boyle:2019kee}],
and the Christodoulou mass \(M_\mathrm{Ch}\)~\cite{Christodoulou:1971pcn}.
The latter is given by
\begin{align} \label{eq : Christodolou mass formula}
    M^2_\mathrm{Ch} \equiv M^{2}_\mathrm{irr} + \dfrac{S^2}{4 M^{2}_\mathrm{irr}},
\end{align}
where \(M^{2}_\mathrm{irr} \equiv A/(16\pi) \), and \(S\) is the magnitude of \(\boldsymbol{S}\).
Ideally, quasi-local mass and spin would be computed using definitions derived in the context of sGB gravity.
However, it remains unclear at the time of writing whether corresponding generalizations have been developed.

For the construction of binary BH initial data, the scale of the system is set by \(M = m_{A} + m_{B}\), where \(m_{A,B}\) are the initial Christodolou masses of the two BHs.
Given an initial separation, orbital velocity and radial velocity, an iterative algorithm controls \(M \to 1\) (in code units) and the spins to the desired values.
In addition, for single BHs, the scalar and driver variables are initialized to zero in both cases, i.e.
\begin{align}
    \Psi \rvert_{t=0} = \partial_t \Psi \rvert_{t=0} &= 0, \nonumber \\
\label{eq:ZeroPsi}
    \boldsymbol{\Sigma} \rvert_{t=0} = \partial_t \boldsymbol{\Sigma} \rvert_{t=0} &= 0.
\end{align}
For the binary, instead, we initialize the scalar field with a small perturbation of the form
\begin{align}
    \Psi \rvert_{t=0} &= 0, & \partial_t \Psi \rvert_{t=0} &= c_0 \alpha (1-\alpha).
\end{align}
where \(\alpha\) is the lapse function computed from the XCTS initial data and \(c_0 / M = 10^{-3}\).
Such a small initial perturbation helps us further test the robustness of the code.
Since this initial data does not correctly describe the scalar field configuration for a binary in quasistationary equilibrium, the system therefore undergoes a transient stage in which it relaxes from the initial configuration to one where both BHs are scalarized.
In general, this also leads to changes on the intrinsic parameters of the binary, including the initial eccentricity and \(m_{A,B}\).

To handle large coupling values, we have also implemented the ramp-up function \(f_\text{ramp}(t)\) of Ref.~\cite{Okounkova:2019zjf} to ameliorate the effects of the transient stage by smoothly turning on the beyond-GR terms.
We introduce this ramp-up function by making the replacement \(\ell^2 \to f_\text{ramp}(t) \, \ell^2\) in Eq.~\eqref{eq : first order system}, where
\begin{align}
  f_\text{ramp}(t) =
  \begin{cases}
    0 & t < t_{s} \\
    F\left(\frac{t-t_s}{t_{\rm ramp}}\right) & t_{s} \leq t \leq t_{s} + t_{\text{ramp}} \\
    1 & t_{s} + t_{\text{ramp}} < t \,.
  \end{cases}
\end{align}
Here, the function
\begin{equation}
     F(x) \equiv x^5  (126 + x (-420 + x (540 + x (-315 + 70 x))))
\end{equation}
takes values \(0\) and \(1\) at \(x=0\) and \(x=1\), respectively, and is chosen such that \(f_{\rm ramp}\) has continuous derivatives up to fourth order.

Eccentricity reduction is carried out using an automatized algorithm designed for GR.
The algorithm fits the time-derivative of the orbital frequency \(\dot{\Omega}\) to an eccentric model inspired by Post-Newtonian (PN) theory in GR.
The fit is performed within a time window \(t \in [t_1, t_2] \), where \(t_1\) is chosen to avoid initial gauge transients, whereas \(t_2\) is chosen to be large enough to encompass \(2-3\) orbits~\cite{Buonanno:2010yk}.
We make sure however, that the window occurs after the ramp up is completed, i.e.~\(t_1 \gtrsim t_s + t_\text{ramp}\).
For equal-mass binaries in scalar Gauss-Bonnet gravity, we find that this algorithm is sufficiently effective to reduce the initial eccentricity to the desired values~\(e \lesssim 10^{-3}\).
Preliminary tests indicate that eccentricity reduction is also effective for some examples of binaries with mass-ratio \(q=2\) and anti-aligned equal-mass spin systems with small-coupling.
It is not guaranteed, however, that this algorithm should work for generic binaries (e.g.~with unequal masses or spins), especially if the inspiral is dipole-dominated, and may thus require an updated PN-inspired model that accounts for changes in the inspiralling rate.
We do not address any possible generalizations here and leave such extensions for future work.


\subsection{Evolution diagnostics and numerical resolution \label{sec : evolution diagnostics}}

As in \paperone, we monitor the effectiveness of the auxiliary variable \(\Sigma_J\) in tracking the target source term \(\mathcal{S}_{J}\).
For this we introduce the following tracking diagnostic
\begin{align} \label{eq : tracking diagnostics}
    \mathcal{E}[\Sigma_{J}](t) \equiv \dfrac{ \lVert \Sigma_J - \mathcal{S}_J \rVert}{\lVert \mathcal{S}_J \rVert + \epsilon},
\end{align}
where \(\Sigma_J \in \{\Sigma, \Sigma_{ab}\}\) and \(\mathcal{S}_{J} \in \{\mathcal{S}, \mathcal{S}_{ab}\}\) is the corresponding source term, \(\epsilon\) is a small number, and \(\lVert \cdot \rVert\) is the \(L_{2}\)-norm computed with the values of the field  components at time \(t\) on all grid-points in the spatial domain.

Additionally, we monitor the constraints arising from the first-order reduction of the scalar system [Eq.~\eqref{eq : first order reduction constraints} in App.~\eqref{app : details first order system}],
as well as the constraint energy \(\mathcal{E}_c\).
The latter is a composite measure that includes several constraints of the first order generalized harmonic system, such as the gauge constraint~\eqref{eq : gauge constraint}
---see Eq.~(53) of Ref.~\cite{Lindblom:2005qh} for the full expression.

For convergence tests we adjust to numerical resolution primarily by adjusting the polynomial order \(p\) of the basis for each element of the spatial domain (\(p\)-refinement).
We label resolutions as \(\mathrm{Lev}n\) (often abbreviated here as \(\mathrm{L}n\)) corresponding to \(p = p_0 + n \),
where \(p_0\) are positive integers chosen empirically for each element and for direction within.


\subsection{Waveform and scalar charge extraction \label{sec : waveform and scalar charge extraction}}

The values of \(\{g_{ab}, \Psi\}\) are recorded during the evolution on a set of extraction spheres \(S_R\) with radii \(R / M \in \{120, 170, 220, 270\}\).
Next, we obtain the GW strain \(h = h_{+} - \mathrm{i} h_{\times}\) at future null-infinity \(\mathcal{I}^{+}\)~\cite{Moxon:2021gbv} by performing Cauchy Characteristic Evolution (CCE).
The scalar wave at \(\mathcal{I}^{+}\) is obtained using the CCE extension for the Einstein-Klein-Gordon system~\cite{Ma:2024bed}.
In both cases, 
the CCE codes propagate the waves to future null infinity by evolving the leading order [\(\mathcal{O}(\ell^0)\)] equations of motion, i.e.
\begin{align}
    R_{ab} = 0, && \Box \Psi = 0.
\end{align}
While we can easily include the canonical stress-energy tensor of the scalar field, as in Ref.~\cite{Ma:2024bed}, here we choose not to do so to keep our GW analysis entirely within the standard CCE and BMS frame-fixing techniques.

The waveforms are rescaled with respect to a reference mass \(M_{\mathrm{ref}}\)
after both the initial intrinsic parameter transients have occurred and the initial burst of junk radiation has left the numerical domain.
We make use of the residual gauge freedom to transform the waveforms to the Bondi-Metzner-Sachs (BMS) \emph{superrest} frame of the binary~\cite{Mitman:2021xkq, Mitman:2022kwt}.
For waveform comparisons, we define an alignment procedure analogous to the \texttt{align2d} method implemented in the \textsc{sxs} package~\cite{SXSPackage_v2025.0.25}.
We align the \((\ell=2, m=2)\)-modes of waveforms \(h\) and \(g\) by minimizing the cost function
\begin{align}
    \mathcal{K}[h, g] \equiv \dfrac{
        \int_{t_{1}}^{t_{2}} dt \, \vert h^{(2,2)}(t + \Delta t)e^{\mathrm{i}\Delta \Phi} - g^{(2,2)}(t) \rvert^{2}
        }{
            \int_{t_{1}}^{t_{2}} dt \, \vert g^{(2,2)}(t) \rvert^{2}
        },
\end{align}
in time and phase shifts \((\Delta t, \Delta \Phi)\).
This function is computed in a time window \([t_{1}, t_{2}]\). The lower bound is chosen such that \(t_{1} / M_{\mathrm{ref}} = 1000 \) is long enough to discard the part of the waveform affected by junk radiation; the upper bound \(t_{2}\) is large enough so that the interval spans \(\sim 3\) orbits, as estimated from the \((2,2)\)-frequency at \(t_{1}\).

The \emph{total} scalar charge of the system can be computed as
\begin{align} \label{eq : scalar charge}
    Q_{R} \equiv -  \dfrac{1}{4 \pi} \oint_{S_{R}} dS \, \hat{s}^{i} \partial_{i} \Psi,
\end{align}
where \(\hat{s}^{i}\) is the outward normal to \(S_{R}\) ---Eq.~\eqref{eq : scalar charge} aims to capture the asymptotic \((1/r)\)-falloff of the scalar field, i.e.~\(\Psi(r \to \infty) = \Psi_{\infty} + Q / r + \mathcal{O}(r^{-2})\), where \(Q \equiv Q_{R\to\infty}\).
For shift-symmetric theories (the example considered here), the scalar charge can be recast in terms of horizon integrals for the component BHs~\cite{Saravani:2019xwx}.
We nevertheless choose to use Eq.~\eqref{eq : scalar charge} to keep our methods valid for other choices of \(f(\Psi)\) that may break shift-symmetry.

Finally, the energy loss in gravitational [c.f.~Eq.~(2.8) of Ref~\cite{Ruiz:2007yx}] and scalar waves can be estimated through the flux-formulas
\begin{align} \label{eq : flux formulas}
    \dot{E}_{\mathrm{gw}} &= \dfrac{1}{2 \kappa }\oint_{S} dS \, \lvert \dot{h} \rvert^2 , 
    &\text{and}& &
    \dot{E}_{\Psi} &= \oint_{S} dS \, \lvert \dot{\Psi} \rvert^2 ,
\end{align}
where \(S\) is a sphere at \(\mathcal{I}^{+}\).


\section{Results \label{sec :  Results}}

In this section, we first test our implementation with evolutions of single BHs (Sec.~\ref{sec :  Results Single Black Holes}), followed by our main results for inspiraling binary BHs.
Sec.~\ref{sec :  Results binary bbh ID} describes the initial data preparation, with Secs.~\ref{sec :  Results binary bbh CD1} and \ref{sec :  Results binary bbh CD2} presenting the results using the CD1 (pathological) and CD2 (well-behaved) drivers.
The extracted waves are discussed in Sec.~\ref{sec : bbh results extracted waves}.
For all simulations, we define our code units by setting \(\kappa = M = 1\).
To facilitate comparisons with results elsewhere, we keep track of these factors below.


\begin{figure}
    \includegraphics[width=\linewidth]{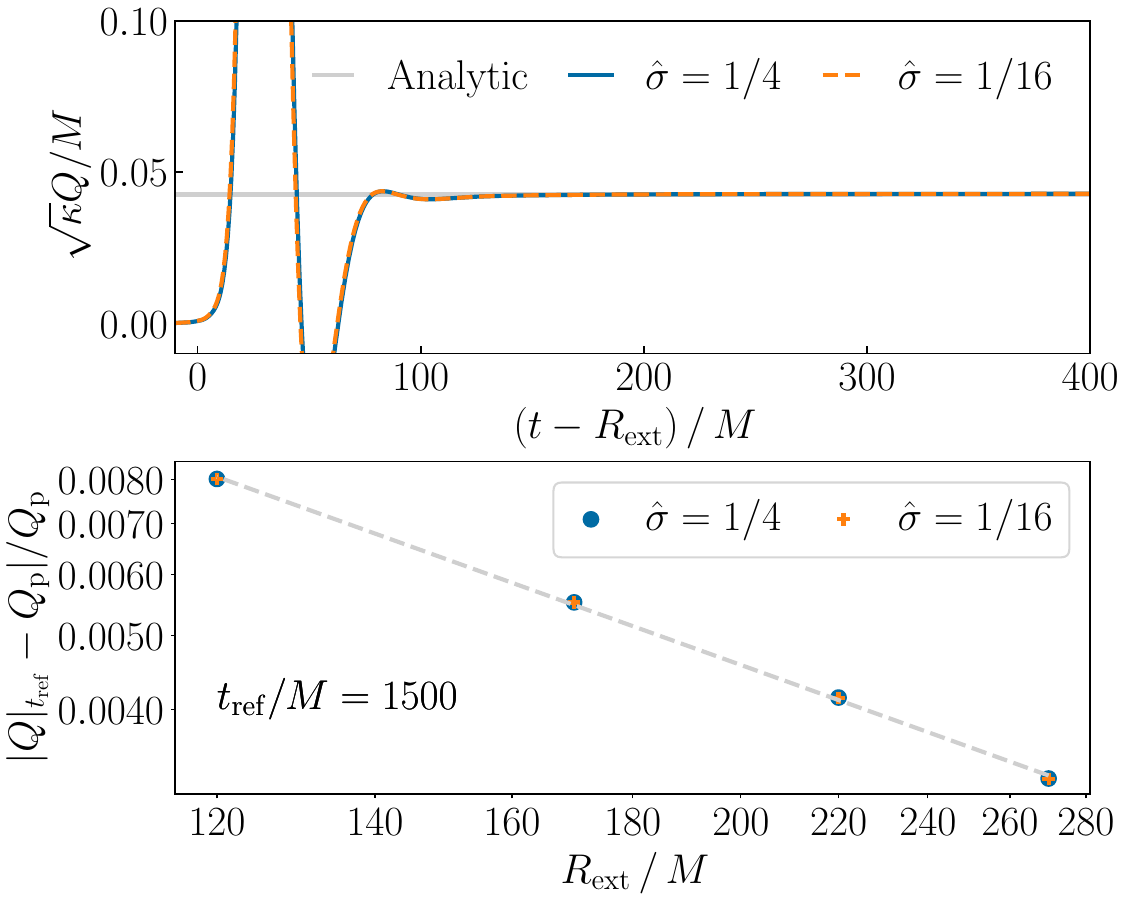}
    \caption{\emph{Scalar charge validation for a spinning BH.}
    Top: Scalar charge [Eq.~\eqref{eq : scalar charge}] in time for a BH with dimensionless spin \(\chi = S/M^2 = 0.67\).
    Bottom: Relative error in the scalar charge when compared to the analytic prediction [Eq.~\eqref{eq : scalar charge analytic prediction}] as a function of the extraction radius.
    For the comparison we choose a reference time \(t / M = 1500\).
    A linear fit (dashed) to the relative error gives \(e^{0.49} \times (M/R_{\mathrm{ext}})^{1.11}\).
    }
    \label{fig : single bh scalar charge validation}
\end{figure}

\subsection{Single black holes \label{sec :  Results Single Black Holes}}

We begin by evolving single, spinning BHs in sGB gravity.  As initial data, we choose the Kerr solution with dimensionless-spin \(\chi = S / M^2 = 0.67 \) along the \(z\)-direction and zero linear momentum.
(Note that these parameters are similar to what we expect for the endstate of an equal-mass non-spinning quasicircular binary.)
We set the initial scalar field to zero, cf.~Eq.~\eqref{eq:ZeroPsi}, and choose a beyond-GR coupling scale of \(\hat{\ell}^2 \equiv \sqrt{\kappa} \ell^2 / M^2 = 1/40\).
Since the linear momentum is zero, the frame-velocity is also zero and both drivers CD1 and CD2 [Eqs.~\eqref{eq : comoving driver simplified} and \eqref{eq : comoving driver simplified second version}] are equivalent.

At the beginning of the simulation, the BH undergoes a transient phase during which the scalar field \(\Psi\) grows, the spacetime geometry adjusts and the system settles into a stationary solution of the theory, which is a scalarized BH.
In the top panel of Fig.~\ref{fig : single bh scalar charge validation}, we show (top panel) the scalar charge \(Q\) obtained at an extraction sphere with radius \(R/M = 170 \) for two values of \(\hat{\sigma}\in \{\hat{\sigma}_1 \equiv 1/4, \hat{\sigma}_2 \equiv 1/16\}\).
Here, we have defined the dimensionless driver parameter \(\hat{\sigma} \equiv \sigma / M^2\).
After an initial pulse, \(Q\) approaches a constant value.

To validate our results, we compare the charge with the analytic approximation \(Q = Q_\mathrm{p} + \mathrm{O}(\ell^4)\) of Refs.~\cite{Yunes:2016jcc, Berti:2018cxi, Prabhu:2018aun}, where
\begin{align} \label{eq : scalar charge analytic prediction}
    Q_\mathrm{p} \equiv 8 \ell^2 \dfrac{S^2 - M^4 + M^{2} \sqrt{M^4 - S^2}}{2 M S^2}.
\end{align}
We compute the analytic prediction using values for the Christodolou mass and spin \((M_\mathrm{ref} \equiv M \vert_{t=t_\mathrm{ref}}, S_\mathrm{ref}  \equiv S \vert_{t=t_\mathrm{ref}})\) after the BH has settled and evaluated at the reference time \(t_\mathrm{ref} / M = 1500 \).
The error in the charge (bottom panel of Fig.~\ref{fig : single bh scalar charge validation}) is dominated by the finite radius extraction and not by the timescales of the \emph{fixing-the-equations} scheme.

\begin{figure}
    \includegraphics[width=\linewidth]{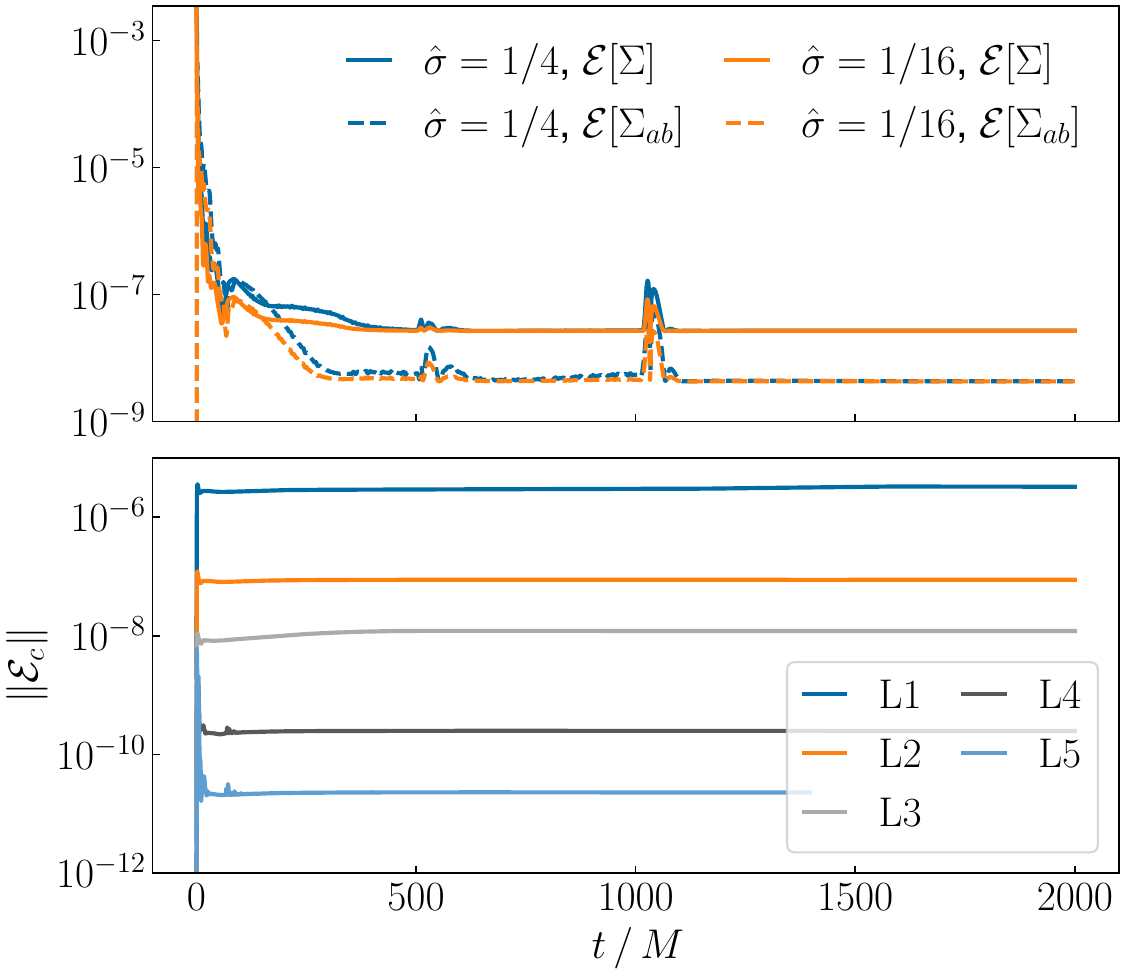}
    \caption{\emph{Single black hole diagnostics in time.}
    Top: tracking diagnostics in time for two different values of \(\hat{\sigma}\in \{\hat{\sigma}_1 \equiv 1/4, \hat{\sigma}_2 \equiv 1/16\}\).
    Solid lines correspond to the scalar auxiliary variables, whereas dashed lines correspond to the tensor variables.
    Bottom: Constraint energy at different resolutions, corresponding to different levels of \(p\)-refinement for the \(\hat{\sigma}_2\)-case.
    }
    \label{fig : single bh tracking in time}
\end{figure}

We now quantify how well the auxiliary variables \(\{\Sigma, \Sigma_{ab}\}\) introduced in our \emph{fixing-the-equations} implementation track the source terms.
In Fig.~\ref{fig : single bh tracking in time}, we focus on the dependence on \(\hat{\sigma}\) at fixed resolution, \(\mathrm{L}3\).
At the start of the simulation \(t/M \lesssim 300\), the magnitude of the diagnostics is larger since the system has yet to settle to the stationary solution. As the system relaxes, these diagnostics decrease in time.
During the initial evolution, smaller driving timescales  (smaller \(\hat{\sigma}\)) enforce a closer tracking of the source terms, which is consistent with our expectations.
Once the system reaches stationarity (\(t/M \gtrsim 300\)), the diagnostics corresponding to the different \(\hat{\sigma}\)-cases approximate the same value and become indistinguishable.
This is expected since the comoving drivers are built to recover the stationary solution exactly, independently of \(\hat{\sigma}\).
(This is also the case for the scalar charge in Fig.~\ref{fig : single bh scalar charge validation}.)
The simulation is stable until the end of each run (at \(t_\mathrm{final}/M = 2000\)), as can be seen from the convergence of the constraint energy \(\mathcal{E}_c\) with resolution for the \(\hat{\sigma}_2\) case (bottom panel of Fig.~\ref{fig : single bh tracking in time}).

\begin{figure}
    \includegraphics[width=\linewidth]{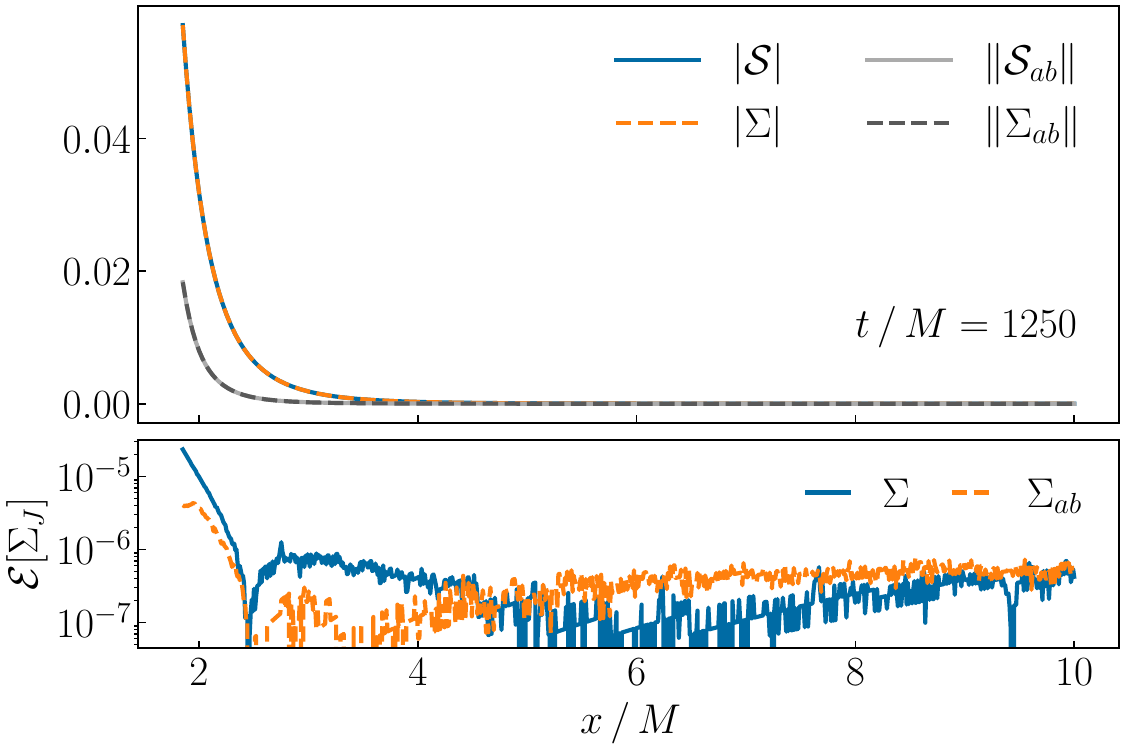}
    \caption{\emph{Single black hole diagnostics along the \(x\)-direction at late time.}
    Top: Spatial dependence of \(\boldsymbol{\Sigma}\) and \(\boldsymbol{S}\) along the \(x\)-direction at time \(t / M = 1250 \).
    Bottom: Tracking diagnostic corresponding to the top panel, which is essentially the relative error.
    }
    \label{fig : single bh tracking radial}
\end{figure}

Now let us examine the spatial configuration of the auxiliary fields in more detail.
In Fig.~\ref{fig : single bh tracking radial}, we show (top panel) the dependence along the \(x\)-direction of both the auxiliary fields and the source terms at a late time \(t/M = 1250\).
The auxiliary fields indeed reproduce the steep gradients in the source terms, including in the region close and within the BH horizon.
The relative error (bottom panel) is \(\lesssim 10^{-4}\) for the \(\hat{\sigma}_2\) case.

\begin{figure}
    \includegraphics[width=\linewidth]{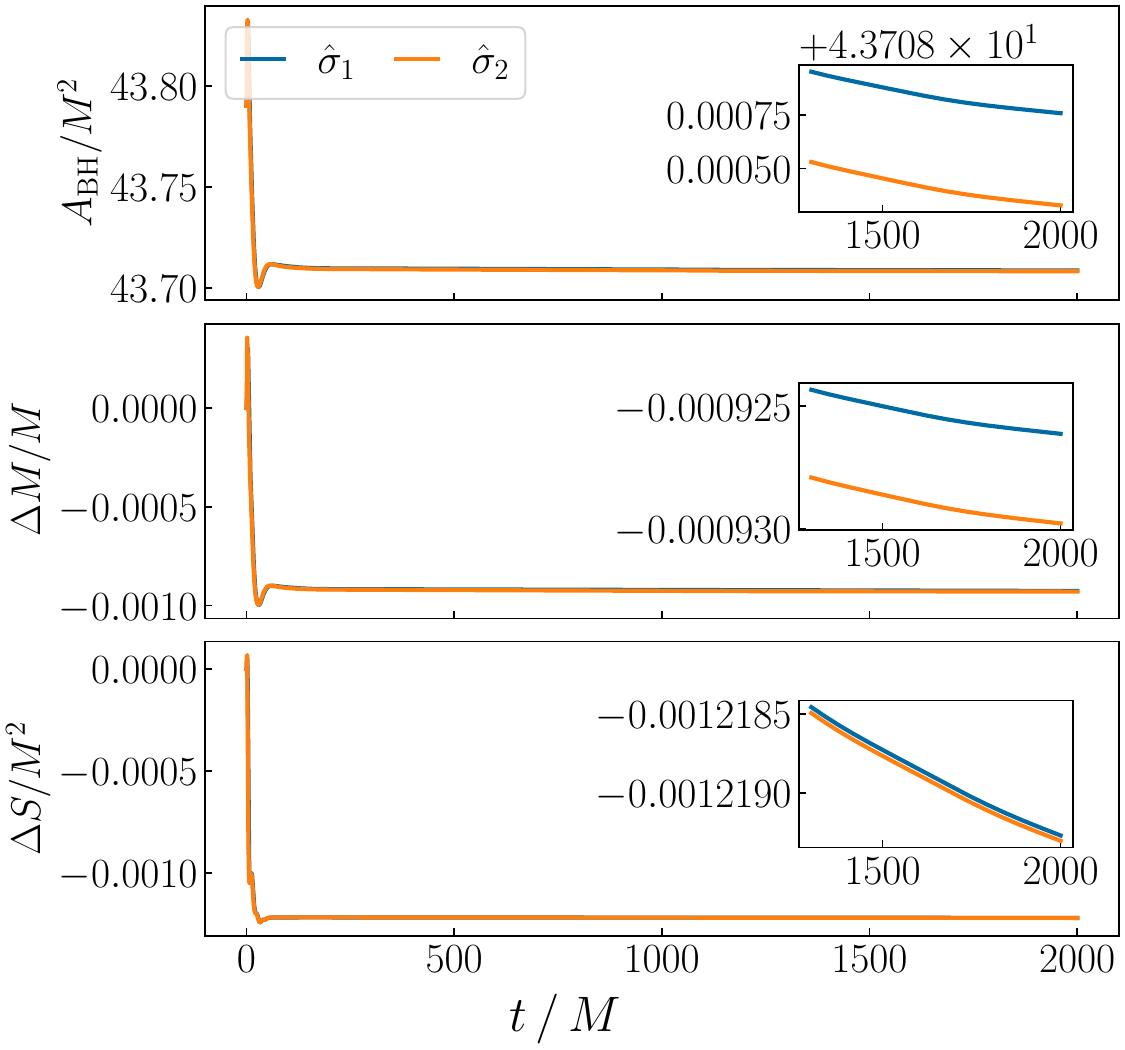}
    \caption{\emph{Single black hole evolutions: Time dependence of black hole
      quantities.}  We show the time dependence of the apparent
      horizon area (top panel), change in mass (middle panel) and
      change in spin (bottom panel). Here we have defined \(\Delta M
      \equiv M_\mathrm{Ch} - M \) and \(\Delta S \equiv S - S
      \vert_{t=0}\), where \(M_\mathrm{Ch}\) is the Christodolou mass
      and \(M_\mathrm{Ch}\vert_{t=0} = M\).
    }
    \label{fig : single bh area mass spin in time}
\end{figure}

Before moving on to the next section, let us remark that an important consequence of the initial transient phase is that it changes the values of the intrinsic parameters of the system;
for binaries, this will affect the orbital parameters, including the eccentricity.
In Fig.~\ref{fig : single bh area mass spin in time}, we plot the changes in area, Christodolou mass and spin for a single BH, as computed from quasi-local formulae (see Sec.~\ref{sec : initial data ramp up and eccentricity reduction}).
Notably, the change in the BH area \(A_\mathrm{BH}\) is \emph{negative}.
(This was also observed in Ref.~\cite{Ripley:2019irj}, and was attributed to the Null Convergence Condition failing to hold for this theory.)
For this case, the Christodolou mass decreases by \(\Delta M / M \approx- 1\%\), where \(\Delta M \equiv M_\mathrm{Ch} - M \);
and the normalized coupling constant increases by \(\Delta(\sqrt{\kappa} \ell^2/M^2) \approx -2 \Delta M / M \approx 2 \%\).


\subsection{Binary black hole initial data preparation \label{sec :  Results binary bbh ID}}

Let us briefly describe how we set up the evolution runs to ensure that we target the desired systems: equal-mass non-spinning quasicircular BH binaries.
The objective is to make our comparisons more robust when comparing evolutions at different resolutions and with different the driver timescales.
This is particularly relevant because, by using initial data in GR (Sec.~\ref{sec : initial data ramp up and eccentricity reduction}), the systems experience a transient phase in which the BHs scalarize (being initially not scalarized), and which causes changes in the initial intrinsic parameters of the system (e.g.~mass, spin, eccentricity).
In the following, we will consider 3 numerical resolutions: L0 (low), L1 (middle) and L2 (high), as well as several driver timescales for the dimensionless parameter \(\hat{\sigma} \equiv \sigma /m^2\), where \(m\) is the Christodolou mass of one of the component BHs.

\begin{figure}
    \includegraphics[width=\linewidth]{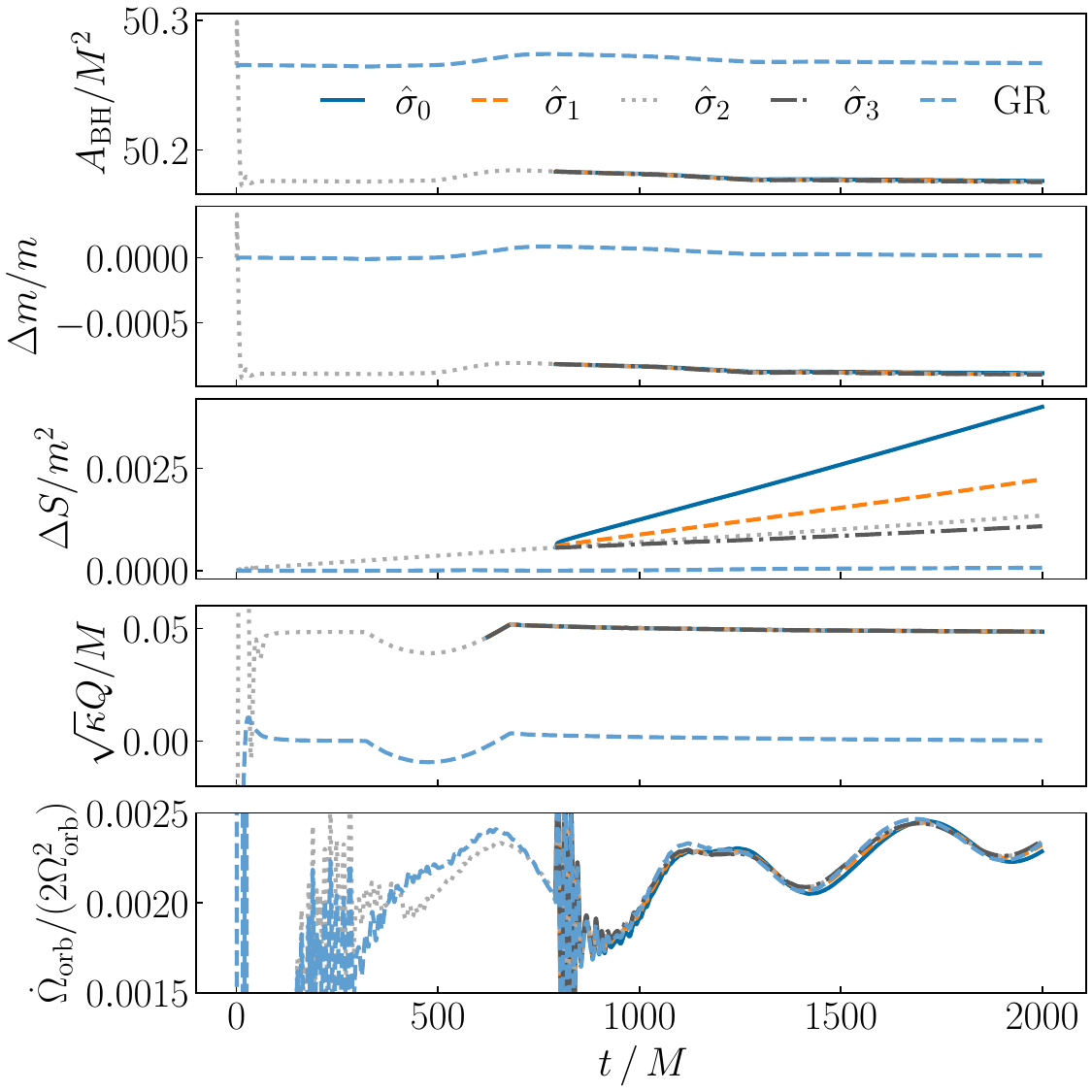}
    \caption{\emph{Binary black hole inspiral:  Time dependence of black hole and orbital quantities (CD1 driver).}
    We show the time dependence of the apparent horizon area (first panel), change in mass (second panel) and change in spin (third panel) during the first few orbits of an evolution using the CD1 driver [Eq.~\eqref{eq : comoving driver simplified}].
    Here we have defined \(\Delta m \equiv m_\mathrm{Ch} - m \) and \(\Delta S \equiv S - S \vert_{t=0}\), where \(m_\mathrm{Ch}\) is the Christodolou mass of a component BH and \(m_\mathrm{Ch}\vert_{t=0} = m\).
    We compare with a simulation in GR (dashed), with the same resolution settings \(\mathrm{L1}\).
    At \(t/M \approx 800\), the \(\hat{\sigma}_{0,1,3}\) runs branch out from the \(\hat{\sigma}_2\) high resolution run,
    where \(\{\hat{\sigma}_0 \equiv 1, \hat{\sigma}_1 \equiv 1/4, \hat{\sigma}_2 \equiv 1/16, \hat{\sigma}_3 \equiv 1/32\}\).
    The total scalar charge of the binary extracted at \(R / M = 170\) is shown in the fourth panel.
    In the bottom panel, the evolution of the orbital frequency \(\Omega_{\mathrm{orb}}\) is shown.
    The amplitude of the oscillations in \(\dot{\Omega}_{\mathrm{orb}} / (2 \Omega_{\mathrm{orb}})\) correspond to the eccentricity of the system, \(e <10^{-3}\).
    }
    \label{fig : binary bh area mass spin in time}
\end{figure}

Initial data \emph{only} is generated for the highest numerical resolution (L2), and lowest (or second-lowest) value of \(\hat{\sigma}\) ---we specify below the actual values used.
A control loop adjusts the initial masses and spins of the XCTS solutions in  GR to the desired values: \(m_{A,B} =1/2\) and \(S_{A,B} = 0\), for BHs A and B ---see Sec.~\ref{sec : initial data ramp up and eccentricity reduction} for more details.
Next, we evolve the fully-coupled sGB system~\eqref{eq : fixed equations of motion} for \(2-3\) orbits, to allow the initial transients to occur (as in previous section) and to obtain enough orbital data to perform a correction to the initial orbital parameters (orbital frequency and radial velocity).
The eccentricity reduction procedure, which involves several iterations of XCTS solves and evolutions is carried out until an eccentricity estimate \(e\) falls below \( 10^{-3}\) ---see also Sec.~\ref{sec : initial data ramp up and eccentricity reduction} for a description of the orbital fitting procedure.
While the eccentricity-reduction procedure ensures that the eccentricity induced by the transient phase is accounted for and corrected,
the change in the other parameters remains (e.g.~the change in the Christodolou masses, i.e.~the total scale of the system); controlling all parameters will require constructing full initial data in sGB ---we will come back to this point in Sec.~\ref{sec : Discussion}.
As an illustration, see Fig.~\ref{fig : binary bh area mass spin in time}, where we show an indicator of the residual eccentricity (bottom panel).
Here, the eccentricity of the system is approximately given by the amplitude of the oscillations in \(\dot{\Omega}_\mathrm{orb}/(2\Omega^{2}_\mathrm{orb})\), where \(\Omega_\mathrm{orb}\) is the orbital frequency obtained from the BH trajectories, and \(\dot{\Omega}_\mathrm{orb}\) is its first time derivative ---see e.g.~Sec.~II of Ref.~\cite{Buonanno:2010yk}.
The other persistent changes in the initial area, mass, spin and scalar charge (shown in the other panels of Fig.~\ref{fig : binary bh area mass spin in time}) will be further discussed in the next two sections, as they are more sensitive to the specific driver equation used.

If the same procedure were to be carried out for different resolutions and different values of  \(\hat{\sigma}\), we expect that such evolutions will be affected by slightly different junk radiation and transients, thus resulting in slightly different physical systems.
To avoid this, we instead record volume data \emph{after} the initial transients have occurred and \emph{after} the main burst of junk radiation has left the spatial domain for the high resolution run on which we carried out the eccentricity reduction procedure.
Subsequent runs are initialized from this data, both at different resolutions and different values of \(\hat{\sigma}\).
We call this procedure ``perform branching after junk'' (abbreviated as ``PBandJ'' or ``pbj'') in \textsc{SpEC} ---see e.g.~Ref.~\cite{Scheel:2025jct}.


\subsection{Binary black holes with scalar drivers \label{sec :  Results binary bbh CD1}}

We now present results for the binary BH problem using scalar comoving drivers for both the scalar \emph{and} tensor auxiliary variables [abbreviated CD1; Eq.~\eqref{eq : comoving driver simplified}].
One of the main conclusions in this section is that the CD1 driver is \emph{not} suitable for long numerical evolutions since the spin magnitude experiences spurious growth during the inspiral ---we will see in the next section that the CD2 driver of the next section avoids this issues.
Nevertheless, we find it illustrative to explore this case in detail.
It will both help us justify the complexity of the CD2 driver (Sec.~\ref{sec : Results binary bbh CD2}), and it is an example of the type of generalization that has been considered so far in the literature.

\subsubsection{Transient phase, convergence and tracking \label{sec : results bbh tracking and transients}}

Consider a system with dimensionless coupling parameter \(\hat{\ell} \equiv \sqrt{\kappa}\ell^2/m^2 = 1/40\) and \(\hat{\sigma} \in \{\hat{\sigma}_0 \equiv 1, \hat{\sigma}_1 \equiv 1/4, \hat{\sigma}_2 \equiv 1/16, \hat{\sigma}_3 \equiv 1/32\}\).
In Fig.~\ref{fig : binary bh area mass spin in time}, we show the area, mass, spin and total scalar charge of the system during the first few orbits.
The change in the mass scale of the system, quantified here by the sum of Christodolou masses, is \(\approx 1 \%\) after the transient phase.
Comparing to a simulation in GR, for which we evolve the system~\eqref{eq : fixed equations of motion} but set (\(\hat{\ell}^2  = 0\)), we conclude that this difference is much larger than that caused by junk radiation in GR.
Note in particular that the values of the mass and scalar charge obtained using different fixing parameters \(\hat{\sigma}\) are indistinguishable from each other (second and fourth panels).
In the corrotating frame, these quantities are \emph{quasistationary} in the early inspiral.
Then, just as in the single BH case, they are well recovered for all values of \(\hat{\sigma}\) due to the effectiveness of our driver equation~\eqref{eq : comoving driver simplified} in reproducing the stationary solutions of the theory.
To more concretely illustrate how effective CD1 is in recovering the BH parameters, we also explicitly compare (App.~\ref{app : appendix comparison with and advection driver}) with results obtained using a driver equation that has different stationary properties.

We expect the spin parameter to also be a quasistationary quantity in the early inspiral. 
However, in the third panel of Fig.~\ref{fig : binary bh area mass spin in time}, we notice that the spin magnitude exhibits approximately linear growth in time.
Empirically, we find that the \(\hat{\sigma}\)-dependence is given by \(d\Delta S /dt \propto \sqrt{\hat{\sigma}}\).
While this may not seem like a large error for the \(\hat{\sigma}_3\)-case, we will see in Sec.~\ref{sec : bbh results larger couplings CD1} that, for larger beyond-GR couplings, the spurious spin-up of the BHs becomes problematic, and we will see in Sec.~\ref{sec : Results binary bbh CD2} that the way to fix this issue is to employ a different generalization of the comoving driver.

\begin{figure}
    \includegraphics[width=\linewidth]{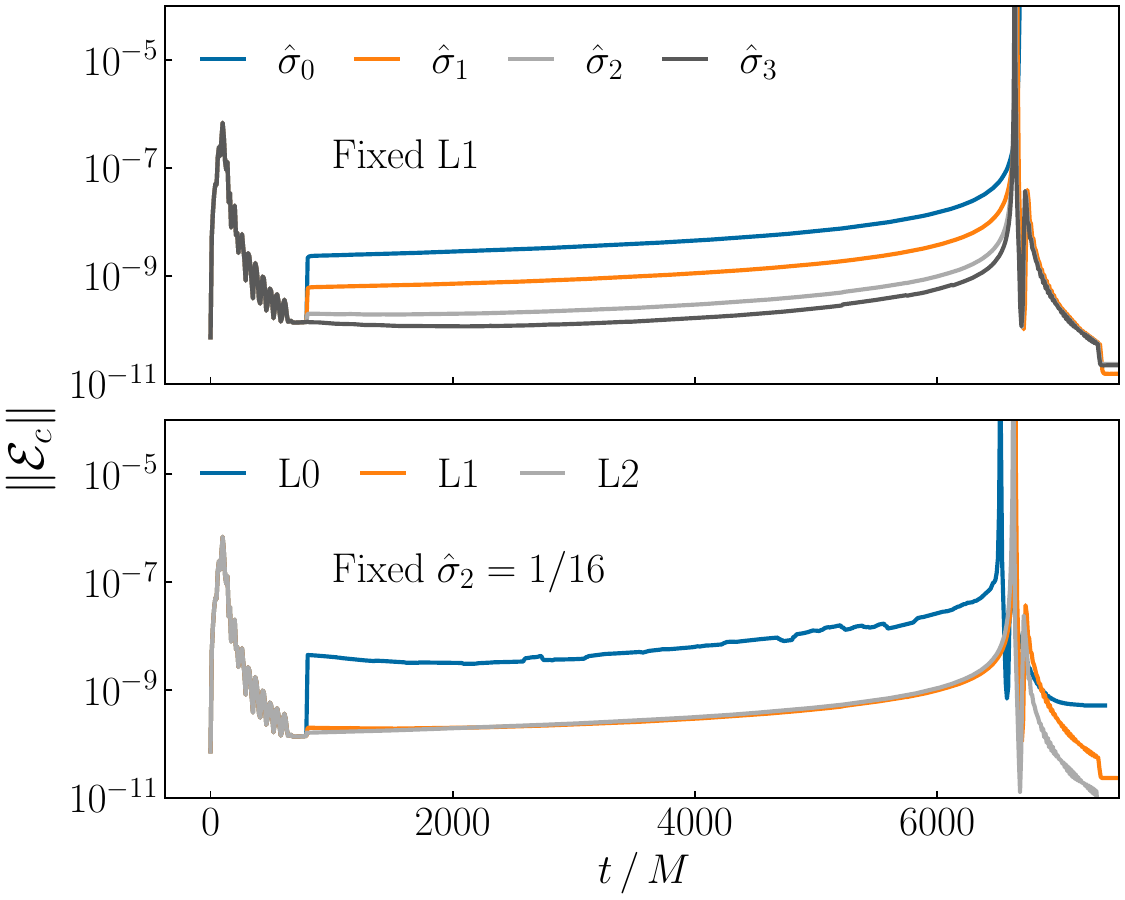}
    \caption{\emph{Constraint energy dependence on resolution and driver timescales.}
    \label{fig : binary constraints}
    Top: \(L_2\)-norm of the constraint energy \(\mathcal{E}_c\) over the entire spatial domain as we keep the resolution fixed to \(\mathrm{Lev}1\) but vary \(\hat{\sigma}\), where
    \(\{\hat{\sigma}_0 \equiv 1, \hat{\sigma}_1 \equiv 1/4, \hat{\sigma}_2 \equiv 1/16, \hat{\sigma}_3 \equiv 1/32\}\).
    At \(t/M \approx 800\), the lower resolution and higher sigma runs branch out from the \(\hat{\sigma}_2\) \(\mathrm{Lev}2\) run.
    Bottom: \(L_2\)-norm of the constraint energy \(\mathcal{E}_c\) as we keep \(\hat{\sigma} = \hat{\sigma}_2\) fixed but vary the resolution (\(p\)-refinement).}
\end{figure}

We are able to evolve the system through merger and ringdown for all cases except for \(\hat{\sigma}_0\), the least effective ``fixing'' parameter.
In Fig.~\ref{fig : binary constraints}, we show the constraint energy \(\mathcal{E}_c\) as we vary both \(\hat{\sigma}\) (top panel) and resolution (bottom panel).
At fixed resolution (\(\mathrm{L}1\)), \(\mathcal{E}_c\) decreases as timescales in the comoving driver are decreased (lower \(\hat{\sigma}\)).
At fixed \(\hat{\sigma} = \hat{\sigma}_2\), \(\mathcal{E}_c\)
decreases when going from the lowest (\(\mathrm{L}0\)) to middle resolution (\(\mathrm{L}1\)), however, it stays roughly the same going from (\(\mathrm{L}1\)) to (\(\mathrm{L}2\)).
From the discussion in Sec.~\ref{sec : gh and constraint propagation}, it is possible that between \(\mathrm{L}1\) and \(\mathrm{L}2\) we are already resolving \emph{non-zero} constraint violations introduced by the \emph{fixing-the-equations} approach.
In order to see \(\mathcal{E}_c\) decrease further at \(\mathrm{L}2\) resolution, we might need to further decrease \(\hat{\sigma}\) below \( \hat{\sigma}_2 \),
however, we do not expect it to be possible to decrease \(\hat{\sigma}\) arbitrarily.

\begin{figure}
    \includegraphics[width=\linewidth]{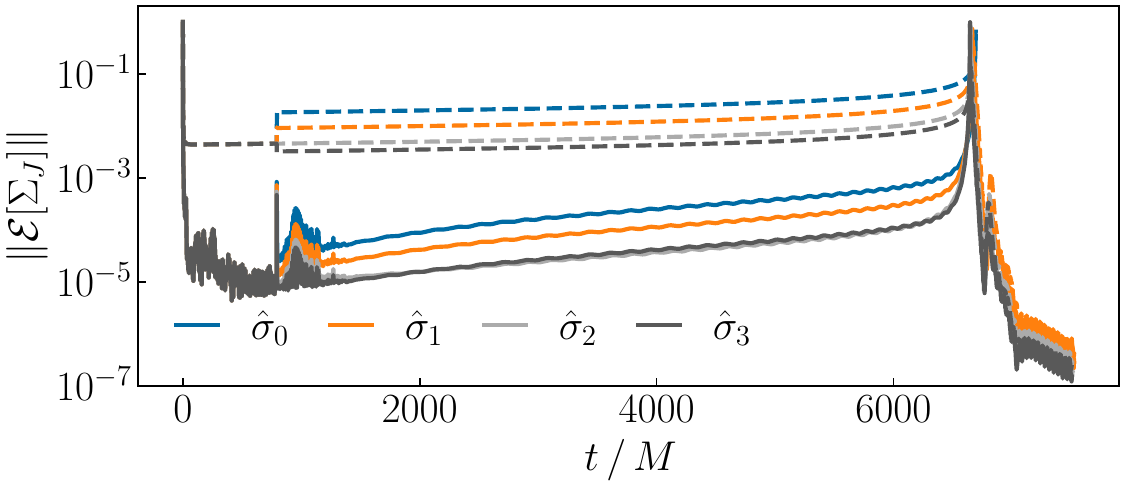}
    \caption{\emph{Tracking diagnostic dependence on driver timescales.}
    \(L_2\)-norm of the scalar tracking diagnostic \(\mathcal{E}[\Sigma]\) (solid) and the tensor tracking diagnostic \(\mathcal{E}[\Sigma_{ab}]\) (dashed) over the entire spatial domain as we keep the resolution fixed to \(\mathrm{Lev}1\) but vary \(\hat{\sigma}\).
    }
    \label{fig : binary tracking}
\end{figure}

The effectiveness of the auxiliary variables \(\boldsymbol{\Sigma}\) in tracking the source terms \(\boldsymbol{\mathcal{S}}\) is controlled by  \(\hat{\sigma}\) is shown in Fig.~\ref{fig : binary tracking} (top panel).
We observe, as expected, that the tracking diagnostics \(\mathcal{E}(\Sigma_{J})\) [Eq.~\eqref{eq : tracking diagnostics}] for both the scalar (solid) and tensor (dashed) variables  decrease with \(\hat{\sigma}\).
In contrast to the single BH case (Sec.~\ref{sec : Results Single Black Holes}), these diagnostic quantities do not saturate at the same value.
Instead, since the system is dynamical during the whole inspiral, we are able to distinguish the dependence on \(\hat{\sigma}\), with lower values corresponding to better tracking.
Here, we can also spot one of the weaknesses of the CD1 driver:
In the top panel of Fig.~\ref{fig : binary tracking}, the tensor variable \(\Sigma_{ab}\) shows tracking diagnostics that are about two orders of magnitude higher than those of the scalar variable \(\Sigma\) during the inspiral.
After merger, the remnant approaches stationarity and both variables exhibit again comparable good performance, as in the single BH case (Sec.~\ref{sec : Results Single Black Holes}).

To further investigate the decreased performance for the tensor variable during the inspiral, we have also performed evolutions for head-on collision systems with identical beyond-GR parameters (not shown here).
For those simulations we find no clear separation in the tracking effectiveness of the scalar and tensor auxiliary variables, indicating that the orbital motion of the system plays a role.


\subsubsection{Larger couplings and spin growth \label{sec : bbh results larger couplings CD1}}

Having quantified the behaviour of the code and of the CD1 driver for a specific small coupling example, we now consider larger values of the beyond-GR coupling and investigate the spin growth issue in more detail.

For larger coupling simulations, we find that we cannot set \(\hat{\sigma}\) arbitrarily low and still retain stability in the numerical evolution.
For \(\hat{\ell}^2 = \hat{\ell}^2_1 = 1/40\), we could set \(\hat{\sigma}\) at least as low as \(\hat{\sigma}_3 = 1/32\).
For \(\hat{\ell}^2 = 2 \hat{\ell}^2_1\), we found however that we need to set \(\hat{\sigma} > \hat{\sigma}_3 \) during the inspiral, whereas for the merger, we need to further increase to \(\hat{\sigma} > \hat{\sigma}_2 =  1/16\).
For \(\hat{\ell}^2 = 3 \hat{\ell}^2_1\), we find that we need \(\hat{\sigma} > \hat{\sigma}_2\) for the inspiral.
However, despite the need for increase values of the driver parameters, these timescales are still \(\sim \mathcal{O}(1 \, M)\), much lower than orbital timescales for the early inspiral.

\begin{figure}
    \includegraphics[width=\linewidth]{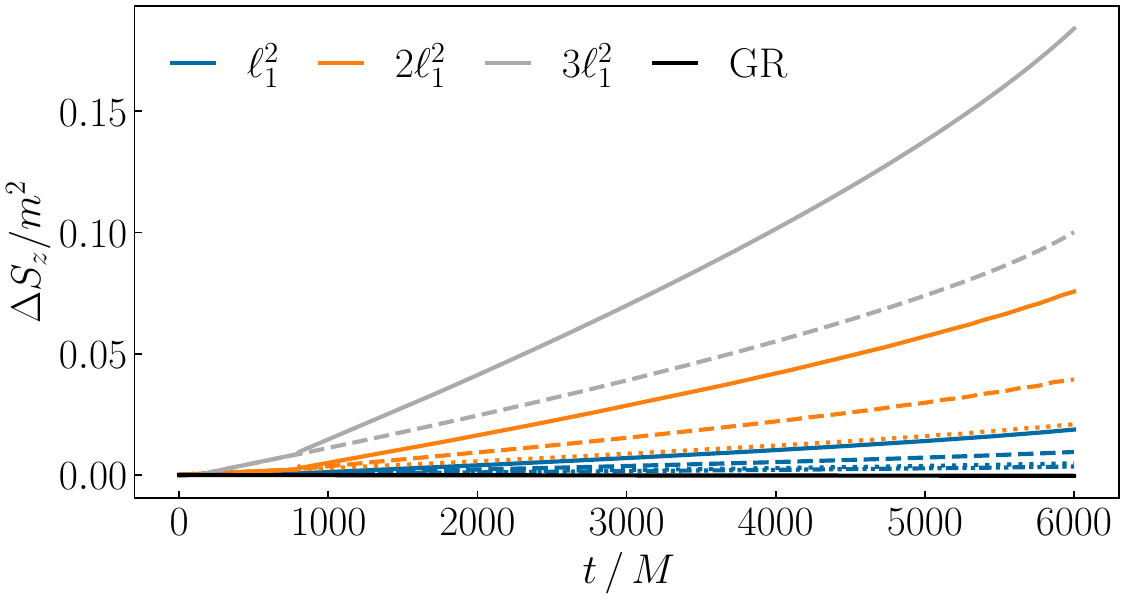}
    \caption{\emph{Aligned spin growth.}
    Growth in the component BH's spin component aligned with the orbital angular momentum for different values of the beyond-GR coupling (color). Different values of \(\hat{\sigma}\) are represented with different linestyles: \(\hat{\sigma}_0=1\) (solid), \(\hat{\sigma}_1=1/4\) (dashed), \(\hat{\sigma}_2=1/16\) (dotted), and, for the  \(\hat{\ell}^2 = \hat{\ell}^1_1\) case, \(\hat{\sigma}_3=1/32\) (dash-dotted).
    }
    \label{fig : larger coupl spin growth}
\end{figure}

For larger couplings, we observe a more rapid spurious growth of the spin component that is aligned with the orbital angular momentum (along the \(z\)-axis in our setup).
In Fig.~\ref{fig : larger coupl spin growth}, one can see that for \(\hat{\ell}^2 = 3\hat{\ell}^2_1\), the spin growth can be as large as \(S/m^2 \sim 0.2 \) at \(t/M = 6000\).
We expect that such serious BH spin-up will lead to a corresponding phase error through orbital hangup (see e.g.~Ref.~\cite{Healy:2018swt}).
This issue seems not to be unique to the CD1 driver, and we also see signs of them in the advective driver explored in App.~\ref{app : appendix comparison with and advection driver}.


\subsection{Binary black holes with the generalized driver \label{sec :  Results binary bbh CD2}}

\begin{figure}
    \includegraphics[width=\linewidth]{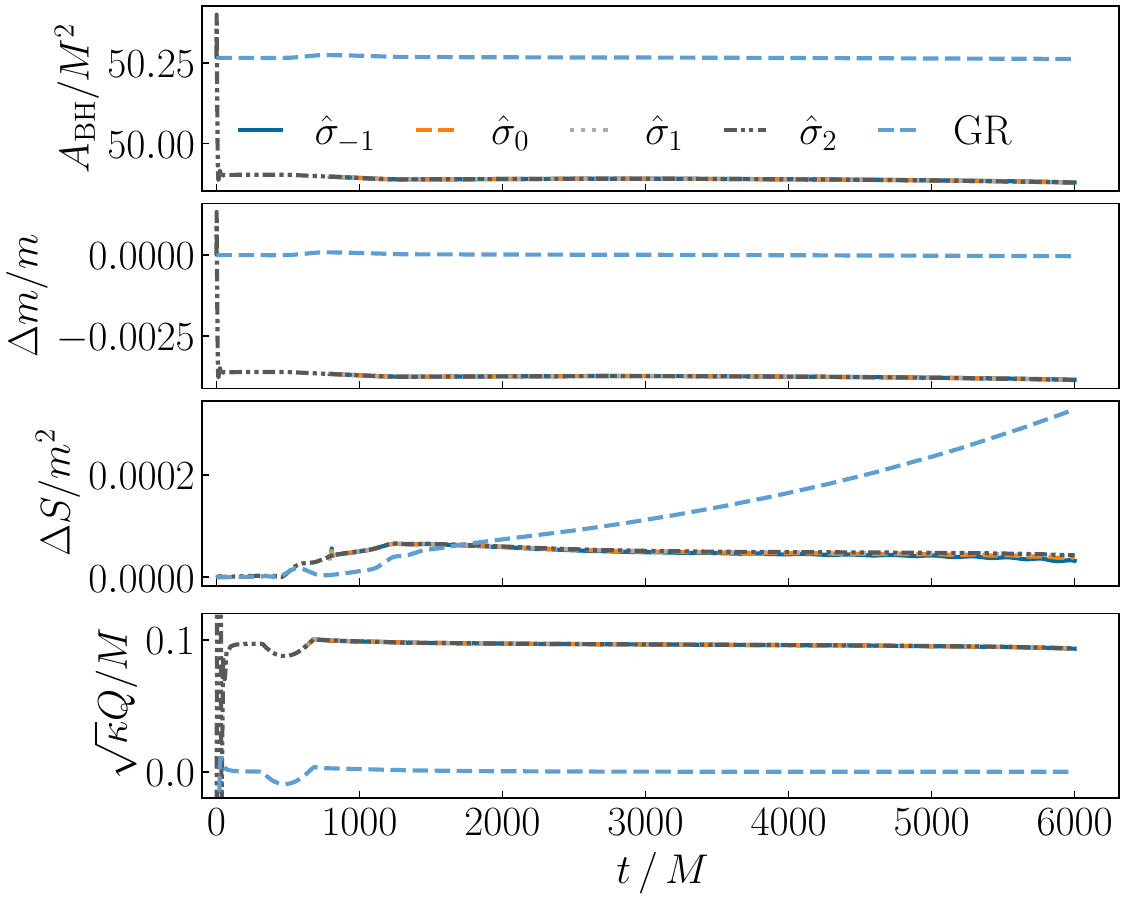}
    \caption{\emph{Time dependence of black hole quantities (CD2 driver).}
    We show the time dependence of the apparent horizon area (first panel), change in mass (second panel) and change in spin (third panel) during the first few orbits of an evolution using the CD2 driver [Eq.~\eqref{eq : comoving driver simplified second version}].
    In contrast to Figs.~\ref{fig : binary bh area mass spin in time} and \ref{fig : larger coupl spin growth}, there is no spin growth during the inspiral.
    The change in spin for the beyond-GR runs, \(\Delta S\), is \(\sim 3\) orders of magnitude smaller than for the corresponding case using the CD1 driver.
    Here, we have defined \(\{\hat{\sigma}_{-1} \equiv 4, \hat{\sigma}_0 \equiv 1, \hat{\sigma}_1 \equiv 1/4, \hat{\sigma}_2 \equiv 1/16\}\).
    }
    \label{fig : binary bh area mass spin in time CD2}
\end{figure}

Evolutions with the CD2 driver [Eq.~\eqref{eq : comoving driver simplified second version}] are \emph{free} of spurious spin growth and of under-performing tensor auxiliary variables \(\Sigma_{ab}\), both issues that where highlighted in the previous section.
In Fig.~\ref{fig : binary bh area mass spin in time CD2}, we show that the CD2 driver recovers the quasistationary behaviour of the BH horizon quantities during the inspiral, including the \emph{spin} for a system with beyond-GR coupling \(\hat{\ell}^2 = 2 \hat{\ell}^2_1 = 1/20 \), or \emph{double} in magnitude with respect our benchmark case of the previous section.
The CD2 generalization avoids any spin growth during the inspiral, with the change in spin magnitude \(\Delta S /m^2\) being \(\sim 3\) orders of magnitude smaller than for the CD1 driver of the previous section (\(\hat{\ell}^2 = 2 \hat{\ell}_1\) case in Fig.~\ref{fig : larger coupl spin growth}).

\begin{figure}
    \includegraphics[width=\linewidth]{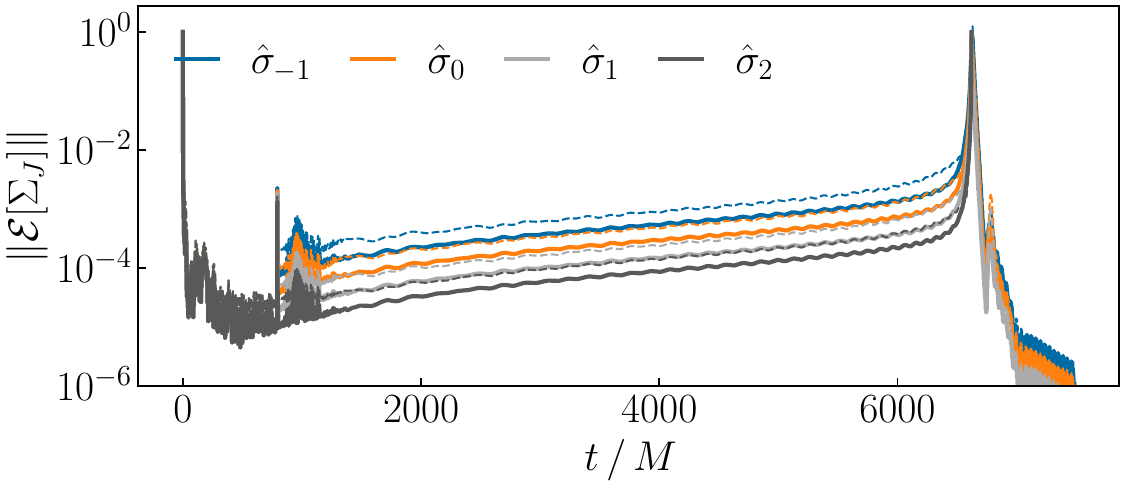}
    \caption{\emph{Tracking diagnostic dependence on driver timescales (CD2 driver).}
    \(L_2\)-norm of the scalar tracking diagnostic \(\mathcal{E}[\Sigma]\) (solid) and the tensor tracking diagnostic \(\mathcal{E}[\Sigma_{ab}]\) (dashed) over the entire spatial domain as we keep the resolution fixed to \(\mathrm{Lev}1\) but vary \(\hat{\sigma}\).
    Here, we have defined \(\{\hat{\sigma}_{-1} \equiv 4, \hat{\sigma}_0 \equiv 1, \hat{\sigma}_1 \equiv 1/4, \hat{\sigma}_2 \equiv 1/16\}\).
    In contrast to Fig.~\ref{fig : binary tracking}, both scalar and tensor diagnostics perform comparatively well.
    }
    \label{fig : binary tracking CD2}
\end{figure}

For this case, the tracking performance of tensor auxiliary variable has also improved by \(\sim 3\) orders of magnitude with respect to the CD1 driver of the previous section (see Fig.~\ref{fig : binary tracking CD2}), bringing it on par with the performance of the scalar variable.
In hindsight, this is due to the CD2 driver taking into account (through a Lie derivative) the frame transformations of the tensor components between the inertial \(\{t, x^{i}\}\) and comoving \(\{\hat{t}=t, x^{\hat{i}}\}\) frames, i.e.~\(
    \Sigma_{ab} (\boldsymbol{x}) \to
    \Sigma_{\hat{a} \hat{b}} (\boldsymbol{x}) = \partial_{\hat{a}} x^{a} \partial_{\hat{b}} x^{a} \Sigma_{ab} (\boldsymbol{x})
\).

\begin{figure}
    \includegraphics[width=\linewidth]{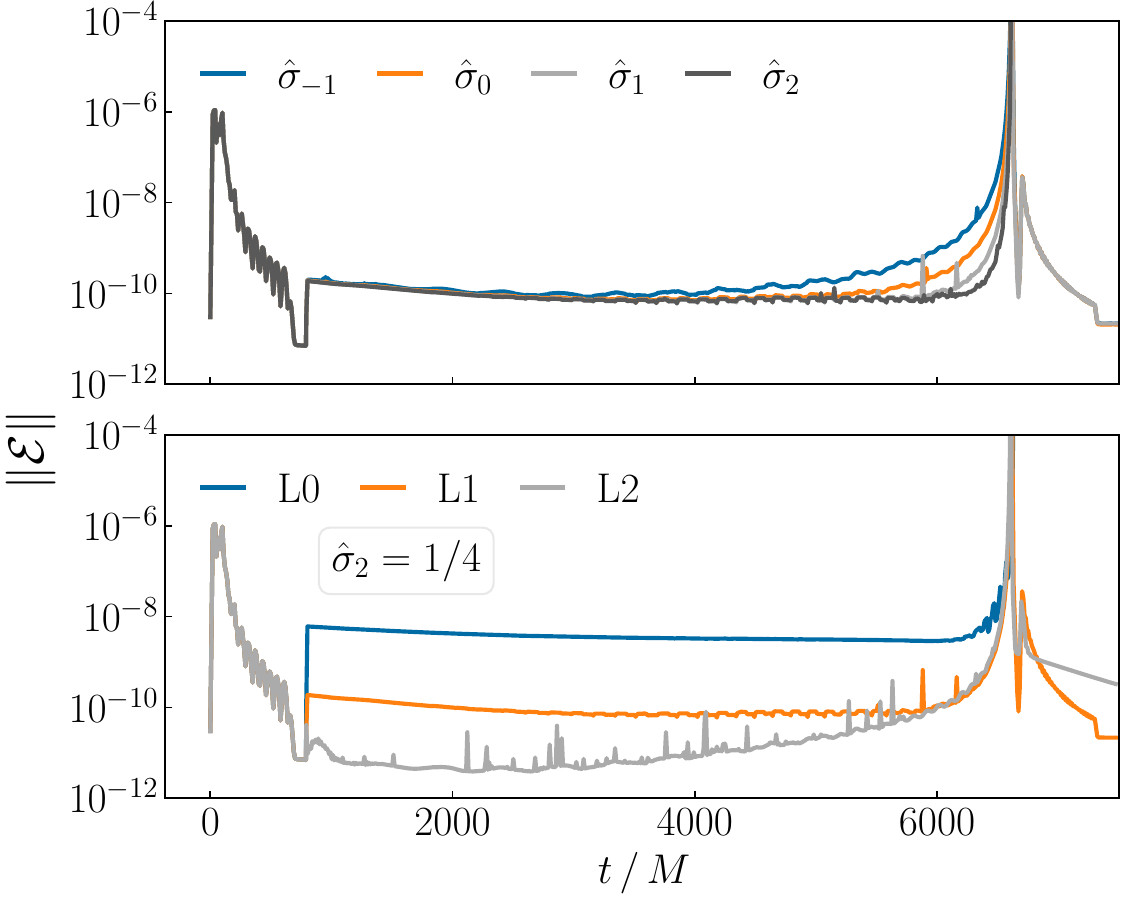}
    \caption{\emph{Constraint energy dependence on resolution and driver timescales (CD2 driver).}
    \label{fig : binary constraints CD2}
    Top: \(L_2\)-norm of the constraint energy \(\mathcal{E}_c\) over the entire spatial domain as we keep the resolution fixed to \(\mathrm{Lev}1\) but vary \(\hat{\sigma}\), where
    \(\{\hat{\sigma}_{-1} \equiv 4, \hat{\sigma}_0 \equiv 1, \hat{\sigma}_1 \equiv 1/4, \hat{\sigma}_2 \equiv 1/16\}\).
    Bottom: \(L_2\)-norm of the constraint energy \(\mathcal{E}_c\) as we keep \(\hat{\sigma} = \hat{\sigma}_1\) fixed but vary the resolution (\(p\)-refinement).}
\end{figure}

We also find a reduction in the non-zero constraint violations, recall the discussion in Sec.~\ref{sec : gh and constraint propagation}, with respect to the previous case.
In Fig.~\ref{fig : binary constraints CD2}, there is a clear separation during the inspiral between our two highest resolutions, with the non-zero constraint violations only noticeable close to merger.

\begin{figure*}
    \includegraphics[width=\textwidth]{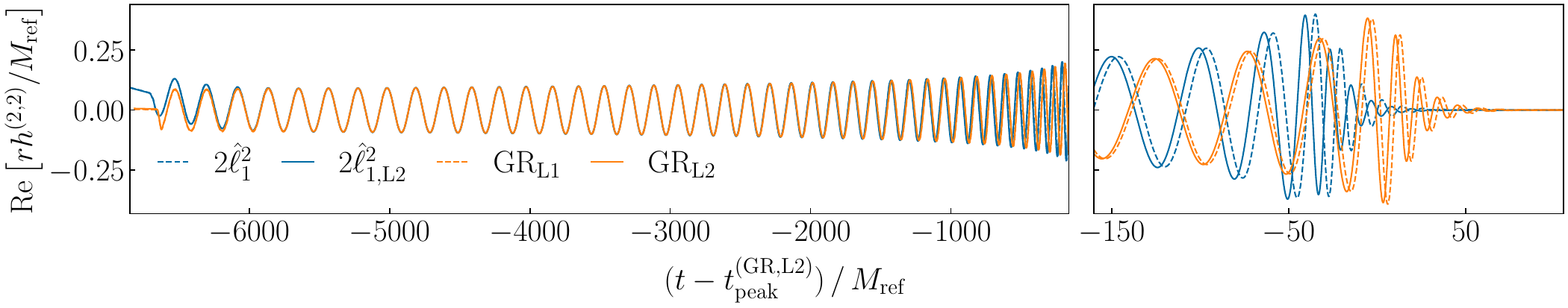}
    \caption{\emph{Merger time comparison for different numerical resolutions.}
    We show the real parts of the GW strain \((2,2)\)-mode extracted with CCE for a simulation with beyond-GR coupling \(\hat{\ell}^2 = 2 \hat{\ell}^2_1 = 1/20\) and for a simulation in GR (\(\hat{\ell}^2 = 0\)).
    The waveform obtained with the highest numerical resolution explored (L2) is shown with solid lines; whereas the middle resolution waveform (L1) is shown with dashed lines.
    All waveforms are aligned to the GR high-resolution waveform in a time window in the early inspiral, and plotted with respect to its peak.
    We find that the merger in the beyond-GR system occurs earlier than in GR.
    }
    \label{fig : merger time difference with GR}
\end{figure*}
\begin{figure*}
    \includegraphics[width=\textwidth]{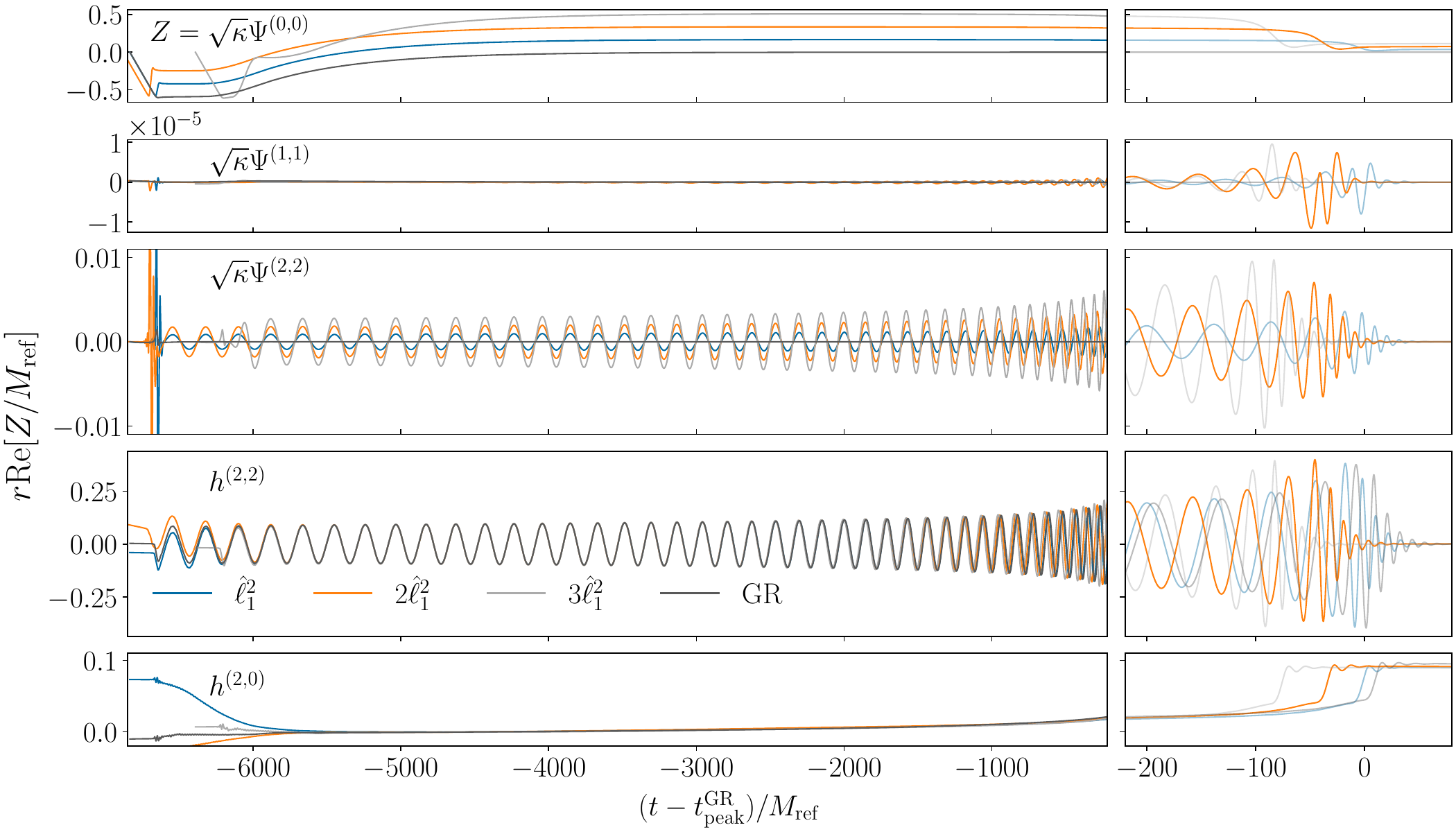}
    \caption{\emph{Gravitational and scalar modes.} 
    We show the real parts of several \((\ell,m)\)-modes for the scalar field (top three panels) nad GW strain (bottom two panels) extracted with CCE; in the right panels we use light colours for all but one case for clarity.
    The waves are mapped to the BMS superrest frame and the time is shown with respect to the GW peak of the \((2,2)\)-mode at \(t_\mathrm{peak}\).
    The dipolar scalar radiation is suppressed for equal-mass non-spinning systems (\(\Psi^{(1,1)}\) in the second panel).
    Memory is visible in the final constant offset of the \(h^{(2,0)}\) mode (bottom panel).
    }
    \label{fig : Showcase gw}
\end{figure*}
\begin{figure}
    \includegraphics[width=\linewidth]{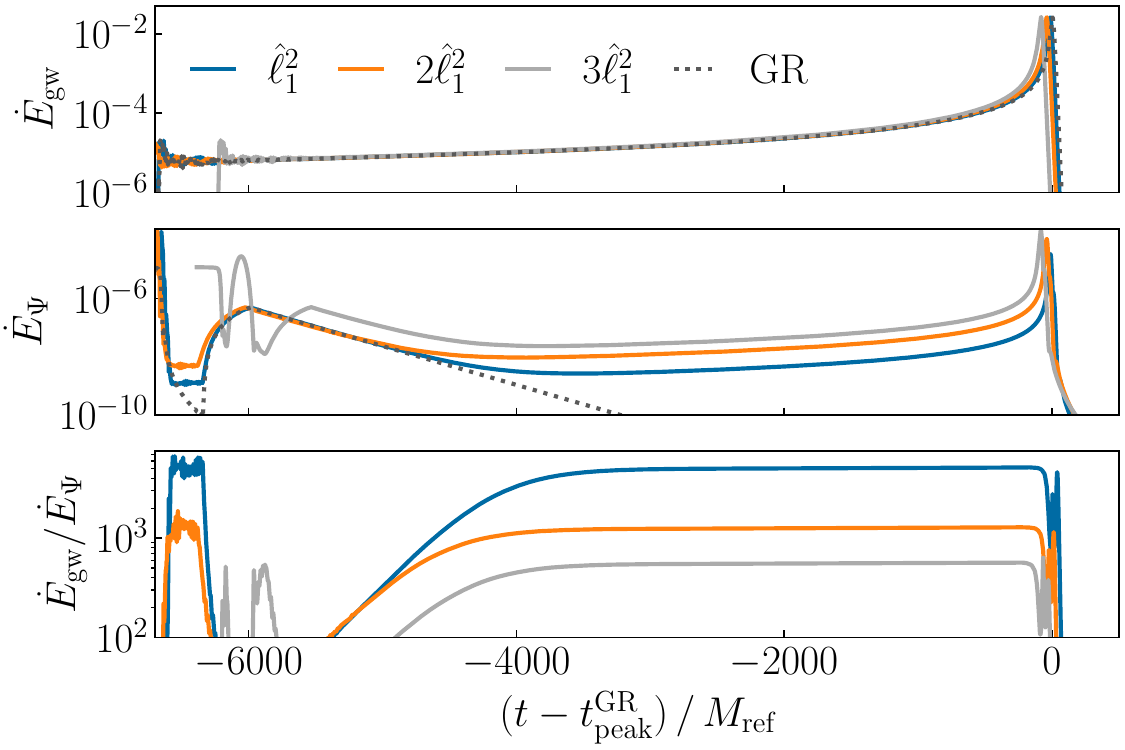}
    \caption{\emph{Energy fluxes.}
    Gravitational (top panel) and scalar (middle panel) energy fluxes, \(\dot{E}_{\mathrm{gw}}\) and \(\dot{E}_{\Psi}\), respectively, for a binary in sGB gravity and in GR.
    The coupling values for the sGB cases are multiples of \(\hat{\ell}^{2}_1 \equiv \sqrt{\kappa}\ell^2 / m^2 = 1/40 \).
    The effects of the initial scalar perturbation and transients are visible in the scalar radiation up to \((t-t^{\mathrm{GR}}_{\mathrm{peak}})/M = 0\),
    where \(t^{\mathrm{GR}}\) is the peak of the GR waveform.
    Bottom: Ratio of the fluxes. The inspiral is clearly dominated by GW energy loss.
    }
    \label{fig : Showcase energy flux}
\end{figure}

\subsubsection{Extracted gravitational and scalar waves\label{sec : bbh results extracted waves}}

With the orbital dynamics under control, we now move on to examine the GW signal.
The GW strain \(h\) is extracted at future null-infinity \(\mathcal{I}^{+}\) using data recorded at \(R/M=170\) for Cauchy Characteristic Evolution (CCE) ---see also Sec.~\ref{sec : waveform and scalar charge extraction} for more details.

To study the dependence of the waveform on \(\hat{\sigma}\) and resolution, we focus on the strain \((2,2)\)-mode.
This mode can be decomposed in amplitude and phase as \(h^{(2, 2)} = A e^{-\mathrm{i}\Phi}\).
To account for the initial transients, once mapped to the superrest frame, we rescale the waveform mass scale by the sum of Christodolou masses \(M_\mathrm{ref}\) computed at the PBandJ time.
In our companion \emph{Letter}~\cite{CompanionLetter}, we presented an analysis of the GW phase error and concluded that
our beyond-GR waveforms are distinguishable from the GR case.
The phase difference with respect to a GR, was found to be \(\sim 3\) orders of magnitude larger than the error due to the \emph{fixing-the-equations} timescales, and \(\sim 10^{-3} \, \mathrm{rad}\).
In Fig.~\ref{fig : merger time difference with GR},
we show the dependence of the real part of \(h^{(2,2)}\) on the numerical resolution, which is the dominant source of phase error.
The waveforms are about 40 GW-cycles long and have been transformed to the inspiral BMS superrest frame; they are then aligned by optimizing on a time window in the early inspiral as described in Sec.~\ref{sec : waveform and scalar charge extraction}.
The phase error in the beyond-GR simulation is comparable to the GR case,
and it is smaller than the overall difference between the two cases,
allowing us to determine that the beyond-GR corrections result in an accelerated merger.

Lastly, we also decompose the scalar field waves in different spherical-harmonic modes \(\Psi^{(\ell, m)}\).
Although dipolar \((\ell=1, m)\) radiation is dominant for generic charged binaries in scalar tensor theories~\cite{Eardley:1975fgi,Damour:1992we} (including sGB gravity), it is suppressed in the case of identical component BHs ---i.e. for equal-mass nonspinning systems.
From Post-Newtonian (PN) calculations (see e.g.~Ref.~\cite{Yagi:2015oca}), we know that the leading-order \(-1\mathrm{PN}\) \((\ell=1, m)\)-modes should vanish due to the fact that both BHs have equal scalar charges, since the corresponding terms are proportional to \((Q_A - Q_B)^2\), where \(Q_{A,B}\) are the scalar charges of the component BHs (A and B).

Using all available modes, we can compute the energy fluxes corresponding to the gravitational and scalar radiation [Eqs.~\eqref{eq : flux formulas}].
In Fig.~\ref{fig : Showcase energy flux}, we show the energy fluxes (top and middle panels), and their ratio (bottom panel), from which we conclude that the energy loss of the evolved system is dominated by GW radiation.


\section{Discussion \label{sec :  Discussion}}

In this section, we elaborate on issues related to the comparability of systems in different theories (Sec.~\ref{sec : discussion system comparison ambiguity}), the limitations of our work (Sec.~\ref{sec : discussion limitations}), and relation to relevant studies in the literature (Sec.~\ref{sec : discussion relation to other work}).


\subsubsection{System comparison ambiguity \label{sec : discussion system comparison ambiguity}}

Changes in the system parameters due to junk radiation and initial transients may be particularly important when comparing observables between GR and beyond-GR, e.g.~when computing phase differences as in our companion \emph{Letter}~\cite{CompanionLetter}.
In our simulations, we have fixed the scale of the system in terms of quasi-local quantities.
Namely, the sum of Christodolou masses at the start of the simulation \(M_\mathrm{Ch}\vert_{t=0} = M = 2m\).
In general, the value of \(M_\mathrm{Ch}\) can change after junk radiation in GR, or after both junk radiation and the initial transient phase in our beyond-GR simulations. (Recall from Sec.~\ref{sec : results bbh tracking and transients} that the effect of the transients is more pronounced than junk.)
For GR, we can correct the waveform (or any other dimensionful quantity) by performing a mass rescaling, in which we use the sum of masses \(M_\mathrm{Ch} \vert_{t=t_\mathrm{relax} }\) at a reference time \(t_\mathrm{relax}\) after junk to normalize the waveform ---see e.g.~App.~A.3.1 of Ref.~\cite{Boyle:2019kee}.
Here, we have chosen the branching time \(t_\mathrm{ref}\), right after the main burst of junk radiation has left the numerical domain.

For sGB, we expect to be able to perform an analogous mass rescaling.
We should keep in mind, however, that it is unclear whether we are justified to apply the general relativistic Christodolou mass formula [Eq.~\eqref{eq : Christodolou mass formula}]~\cite{Christodoulou:1971pcn} for our sGB evolutions.
Therefore, naively normalizing the waveforms using \(M_\mathrm{Ch}\) (which is the procedure adopted in Sec.~\ref{sec : bbh results extracted waves}) is not guaranteed to result in waveforms that are strictly comparable between the GR and sGB.
Entropy formulas (see e.g.~Refs.~\cite{Wald:1993nt, Julie:2019sab, Hollands:2022fkn}) are a useful step forward in identifying the generalization of the irreducible mass,
however, we are unaware of any complete generalizations of the Christodolou formula that would allow for a separation of mass, spin and scalar contributions to the irreducible mass.

One may choose instead to try to match generalized ADM momenta that include scalar corrections and account for the binding energy of the system.
Being global quantities, it may however be difficult to use them to compare generic binaries, as they do no provide information for the individual BH parameters.
Moreover, accurately computing these quantities requires placing the integration surfaces at large radii.
In practice, for GR, we only compute the ADM quantities during the initial data generation stage (for which numerical domain can extend up to \(R_\mathrm{out}/M \gtrsim 10^{5}\)), and do not keep track of them during the evolution (for which the boundary is \(R_{out} / M \gtrsim 500\)).

One other alternative is to compare waveforms purely in terms of observables.
For instance, given waveform in theory \(B\), find the closest matching waveform in the parameter-space of theory \(A\) in a time (frequency) window \(\mathcal{W}\).
This approach would also be consistent with how parameter estimation might work in practice, if full (beyond-GR) waveform models were to be used on observational data.
As an example, consider the quasicircular equal-mass nonspinning case.
Given a fixed waveform \(\hat{h}_{B}\) in sGB with mass scale parameter \(\hat{M}_B\), one could find the optimal GR-waveform \(\hat{h}_{A}\), with mass scale \(\hat{M}_{A}\), that minimizes the mismatch \(\mathcal{MM}\), i.e.
\begin{align}
    \mathcal{MM} \left[\hat{h}_A, \hat{h}_B \right] = \min_{M_A} \mathcal{MM}\left[h_A(M_A), \hat{h}_B\right],
\end{align}
where
\(\mathcal{MM}[h_A, h_B] \equiv 1 - \langle h_A \vert h_B \rangle / \sqrt{\langle h_A \vert h_A \rangle \langle h_B \vert h_B \rangle}\)
and \(\langle \cdot \vert \cdot \rangle\) is and inner product restricted to \(\mathcal{W}\).
The best match would depend, however, on the choice of  \(\mathcal{W}\).
For quasicircular systems, a natural choice of \(\mathcal{W}\) would encompass few orbits during the early inspiral, during which the frequency evolution is slow.


\subsubsection{Limitations \label{sec : discussion limitations}}

Among the main limitations in our work is our use of binary BH initial data constructed in GR for our simulations in sGB, which does not describe an equilibrium configuration, thus leading to transients that alter the initial parameters of the system.
However, constraint-satisfying quasistationary initial data has not yet been constructed for scalarized binaries in this theory ---in Ref.~\cite{Nee:2024bur} we have made progress by solving for quasistationary initial data in the test-field approximation, whereas Ref.~\cite{Brady:2023dgu} considered backreaction given a fixed scalar profile which is not in equilibrium.
Using such data would allow us to perform simulations that better target specific parameters by reducing transients.
It will also reduce the impact of the mitigating strategies to deal with them, such as ramp-up phases, truncation of the waveform, and longer partial evolutions for eccentricity control.
In ongoing efforts, we are attempting to address this problem by fully implementing our recent extension of the XCTS formalism to scalar-tensor theories~\cite{Nee:2024bur}.
Regarding eccentricity control, one would in principle also need to generalize the eccentric fitting model to account for physics beyond GR, especially for cases (if any), where the inspiral rate clearly different from GR, such as in dipole-dominated binaries.
For the purposes of eccentricity reduction, however, we have not seen an immediate need for this.

Here, we have chosen to focus our simulations to cases where the beyond-GR coupling is relatively small and for which we have good control of the evolution.
For equal-mass systems, we have found that cases where \(\sqrt{\kappa}\ell^2/m^2 \gtrsim 3/40 \) (the largest coupling value shown here) start become increasingly more challenging.
This is due to our fixed mesh refinement approach, and the need to use larger values for driver parameters (which are keep fixed throughout the whole simulation).
Indeed, we have found empirical evidence that the driver parameters may be limited by the scale of the beyond-GR coupling, and roughly given by,
\begin{align}
    \sigma \gtrsim \ell^2 ,
\end{align} 
which was observed analytically in toy examples in Ref.~\cite{Cayuso:2017iqc}, and numerically for a spontaneously scalarized BHs in Ref.~\cite{Franchini:2022ukz}.
As discussed in Sec.~\ref{sec : gh and constraint propagation}, and illustrated in Secs.~\ref{sec : results bbh tracking and transients} and \ref{sec :  Results binary bbh CD2}, a minimum allowed value of \(\sigma\) has consequences for the amount of non-zero constraint violations in our simulations, and in theory, limiting their accuracy.
Since the explored cases presented here seem to be mainly limited by numerical resolution (Sec.~\ref{sec : bbh results extracted waves}), this may not be problematic for now. 
Moreover, adaptative mesh refinement (AMR) in future versions of \textsc{spectre} will allow us to use more efficient numerical domains to balance the changing scales of the system (e.g.~separation for eccentricity systems) and to keep the constraints under control during the entire simulation.

Regarding wave propagation, we do not account for modifications of the propagation speeds of scalar and tensor modes arising from the \(\mathcal{O}(\ell^2)\)-terms.
First, for the Cauchy evolution, this approximation is deeply ingrained in our application of the \emph{fixing-the-equations} approach, and arises from moving the terms \(\mathcal{O}(\ell^2)\)-terms in the principal part to the RHS in Eqs.~\eqref{eq : fixed equations of motion}.
And second, for our wave extraction procedure at future null infinity \(\mathcal{I}^{+}\), we have ignored all \(\mathcal{O}(\ell^2)\)-terms altogether, including for the procedure to perform the BMS superrest frame fixing.
The nontrivial terms \(\propto \ell^2 H^{\mathrm{(TR)}}_{ab}\) decay rapidly with distance (\(\sim r^{-4}\)) ---this was argued, e.g.~in Sec.~III.C of Ref.~\cite{Corman:2022xqg}.


\subsubsection{Relation to other work \label{sec : discussion relation to other work}}

\begin{figure}
    \includegraphics[width=\linewidth]{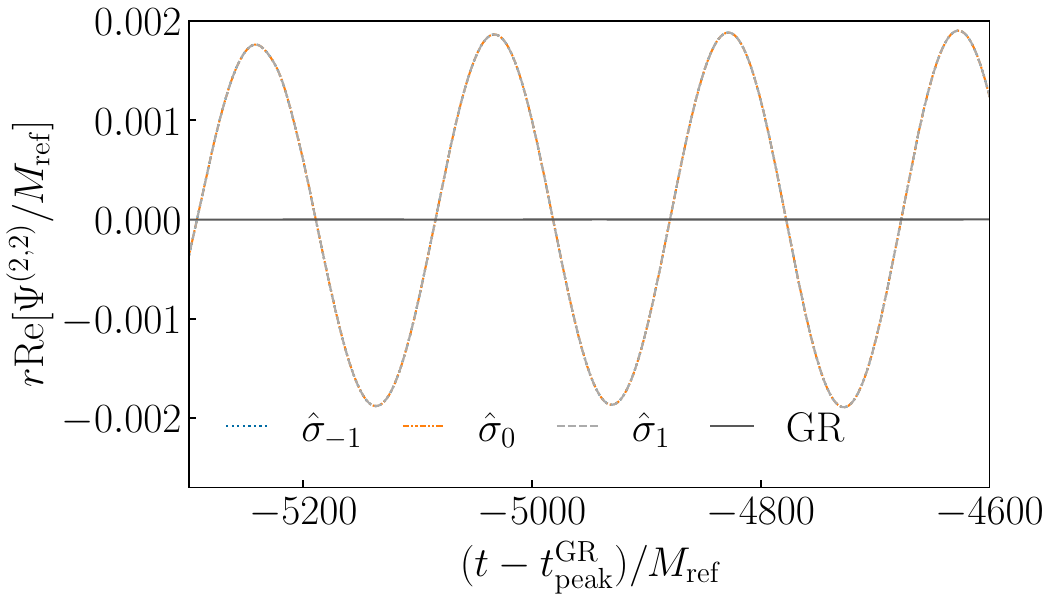}
    \caption{\emph{Inspiral quadrupolar scalar wave dependence on \(\hat{\sigma}\).}
    The quadrupolar scalar wave amplitude is a proxy of the description of the intrinsic parameters of the BH components.
    The quasistationary properties of the CD2 driver [Eq.~\eqref{eq : comoving driver simplified second version}] make the wave amplitude relatively insensitive to the driver parameter \(\hat{\sigma} \in \{\hat{\sigma}_{-1} \equiv 4, \hat{\sigma}_0 \equiv 1, \hat{\sigma}_1 \equiv 1/4, \hat{\sigma}_2 \equiv 1/16\}\).
    In the case of a wavelike driver [Eq.~\eqref{eq : wavelike driver}], or the advection driver of App.~\ref{app : appendix comparison with and advection driver}, the BH intrinsic parameters (e.g.~scalar charge) depend significantly on \(\hat{\sigma}\), which translates into a significant dependence in the scalar wave amplitude.
    This figure is to be contrasted with the right panel of Fig.~14 in Ref.~\cite{Corman:2024cdr}.
    }
    \label{fig : inspiral scalar wave zoom}
\end{figure}

Ref.~\cite{Corman:2024cdr} (hereinafter \paperB) is the only other work that carried out binary evolutions in the shift-symmetric version of sGB using the \emph{fixing-the-equations} approach, mainly with the objective of comparing results between different formulations (the \emph{fixing-the-equations}, modified GH and reduction-of-order systems).
We therefore compare our results against those of \paperB~to better contextualize the improvements achieved here;
while these advances are applicable to other examples using the same method, such as the cases explored in Refs.~\cite{Bezares:2021yek, Cayuso:2023xbc}, it is not easy to identify close parallels to those cases.

We can identify the fields, couplings and driver parameters in \paperB~(labelled with \(\star\) subindices) with ours as
\begin{align} 
    \left\{
    \phi_{\star} = \sqrt{\kappa} \Psi,
    \dfrac{\lambda_{\star}}{m^2} = \sqrt{\kappa} \dfrac{\ell^2}{m^2},
    \dfrac{\sigma_{\star} }{\kappa_{\star} \alpha^2} = \sigma, 
    \kappa_{\star} = \dfrac{1}{\tau} \right\} , \nonumber
\end{align}
with derived timescales
\(
    \{
        t_{1, \star} = 2 \alpha^2 \sigma/\tau, t_{2, \star} = \alpha \sqrt{\sigma}
    \}
\),
where \(\alpha\) is the lapse function and \(\kappa_{\star}\) is also a driver parameter.

Notice however that \paperB~uses a wavelike driver of the form
\begin{align} \label{eq : wavelike driver}
    -\sigma_{\star} g^{ab}\partial_{a} \partial_{b} \boldsymbol{\Sigma} + \partial_{t} \boldsymbol{\Sigma} = - \kappa_{\star} (\boldsymbol{\Sigma} - \boldsymbol{\mathcal{S}}) 
.
\end{align}
The simulations presented in \paperB~explore beyond-GR coupling values \(\sqrt{\kappa} \ell^2/m^2 \in [0.02, 0.25]\) for single BHs, and \(\sqrt{\kappa} \ell^2/m^2 \in [0.02, 0.1]\) for binaries;
our binary simulations fall on the lower end of these ranges \(\sqrt{\kappa} \ell^2/m^2 \in [0.025, 0.75]\).
The driver parameters in \paperB~are roughly \(\hat{\sigma} \in [0.04,  1.1]\) for single BHs, whereas for inspiralling binaries, they are \(\hat{\sigma} \in [5, 36]\);
instead, we explore significantly lower values \(\hat{\sigma} \in [0.03125, 1]\).
However, \paperB~studies binary systems with large mass-ratio \(q = 1.22\) which makes our binary setups not exactly comparable.

While serving the same purpose as the comoving drivers studied in this work, the main qualitative difference in evolutions using the wavelike driver of \paperB~arises from the differences in their stationary limit.
For time-independent fields, the tracking given by wavelike driver reduces roughly to \((\boldsymbol{\Sigma} - \boldsymbol{\mathcal{S}}) \approx -\sigma \nabla^2 \boldsymbol{\Sigma}  \).
And just as CD1 in this work [Eq.~\eqref{eq : comoving driver simplified}], \paperB~treats the tensor auxiliary variables as if they were scalars.
We note that \paperB~did not report any undesired spin growth during the evolution; it is possible that this may not have been apparent to the authors due to the short duration of the simulations, of about 1000 \(M\).

The driver choices in \paperB~\emph{are} theoretically valid according to the \emph{fixing-the-equations} approach, and have been useful in carrying out explorations for the two-body problem in sGB gravity using this method.
However, we have shown here that better choices exist.
Our main proposal, the comoving driver CD2 [Eq.~\eqref{eq : comoving driver simplified second version}] yields significantly improved accuracy and efficiency in such simulations in all respects.
We highlight that obtaining similar accuracy with a wavelike driver would require one to specify prohibitively small values of the driver timescales ---either due to stiffness of the equations or other more fundamental reasons such as those discussed in Sec.~\ref{sec : discussion limitations}.
The main inefficiency in wavelike drivers of the form~\eqref{eq : wavelike driver} lies \emph{primarily} in the poor description of the individual objects, which needs to be overcome by specifying smaller timescales, and \emph{not} in the timescales of the binary problem actually changing for each driver choice.
We refer the reader to App.~\ref{app : appendix comparison with and advection driver}

The improvement in the component BH description is already evident for single BH simulations (Sec.~\ref{sec : Results Single Black Holes}), for which the BH parameters (including the scalar charge) do \emph{not} exhibit such a strong dependence on the \emph{fixing-the-equations} timescales as that shown in Sec.~V A of \paperB (or in Ref.~\cite{Franchini:2022ukz}).
One of our main design contributions is to use comoving derivatives to allow this property to hold for almost the entire inspiral of a BH binary; in the comoving frame, the component BHs behave essentially as in the isolated BH case.
In Fig.~\ref{fig : inspiral scalar wave zoom}, we have zoomed into the quadrupolar scalar wave during the inspiral to show how the quasistationary behaviour of the comoving driver makes the scalar wave relatively insensitive to the driver parameters \(\hat{\sigma}\).
This is in contrast to the right panel of Fig.~14 of \paperB, for which the waveform amplitude shows significant variation with \(\hat{\sigma}\).


\section{Conclusions \label{sec :  Conclusions}}

Here, we have presented in detail our implementation of the \emph{fixing-the-equations} approach~\cite{Cayuso:2017iqc} in \textsc{spectre}, an NR code employing a discontinuous Galerkin scheme~\cite{deppe_2026_19373346}.
The improvements in length and accuracy of the simulations obtained both here and in our companion \emph{Letter}~\cite{CompanionLetter} arise mainly from: \emph{i)} the use of efficient spectral methods; \emph{ii)} our proposal to exploit the approximate symmetries of BH binaries to write accurate comoving drivers [Eq.~\eqref{eq : comoving driver} or Eq.~\eqref{eq : comoving driver simplified second version}] to do the \emph{fixing}; \emph{iii)} the use of other state-of-the art methods in GR to perform wave extraction and eccentricity control.
We also discussed extensively (Sec.~\ref{sec : Results binary bbh CD1}) the importance of treating tensor auxiliary variables properly to avoid underperformance and spurious effects (namely, spin growth), as well as to justify the complexity of Eq.~\eqref{eq : comoving driver}.

Our code can be extended to explore the larger family of sGB gravity and possibly other theories beyond GR (e.g.~EFT extensions of vacuum GR~\cite{Cayuso:2020lca, Cayuso:2023xbc}, \(k\)-essence~\cite{Bezares:2021yek, Lara:2021piy, Coates:2023swo}, vector theories~\cite{Rubio:2024ryv}), once the appropriate source terms in the ``fixed'' system of equations of motion are specified.
Future work will extend the evolutions to the case of unequal mass and spinning cases, and construct complete initial data for sGB binaries.
We will also use these waveforms for waveform model comparisons and other applications regarding tests of GR.
%


\begin{acknowledgments}
    The authors would like to thank Alessandra Buonanno, Fabrizio Corelli, Luis Lehner, Gonzalo Morr\'as, Oliver Markwell, Peter James Nee, Raj Patil, and Sebastian V\"olkel for fruitful discussions.
    Computations were performed on the Urania HPC systems at the Max Planck Computing and Data Facility.
    This material is based upon work supported by the National Science Foundation under Grants No.~PHY-2309211; No.~PHY-2309231;  No.~OAC-2513339 at Caltech; and NASA award No.~80NSSC26K0340, and No.~PHY-2407742; No.~PHY-2207342; No.~OAC-2513338; and NASA award No.~80NSSC26K0340 at Cornell; NSF Grants NSF AST-2219109, PHY-2208014, Nicholas and Lee Begovich, and the Dan Black Family Trust. Any opinions, findings, and conclusions or recommendations expressed in this material are those of the author(s) and do not necessarily reflect the views of the National Science Foundation or NASA. This work was supported by the Sherman Fairchild Foundation at Caltech and Cornell.
    NLV acknowledges support from the Swiss National Science Foundation (SNSF) Ambizione grant PZ00-2\_232961.
\end{acknowledgments}

\FloatBarrier


\appendix


\begin{figure}
    \includegraphics[width=\linewidth]{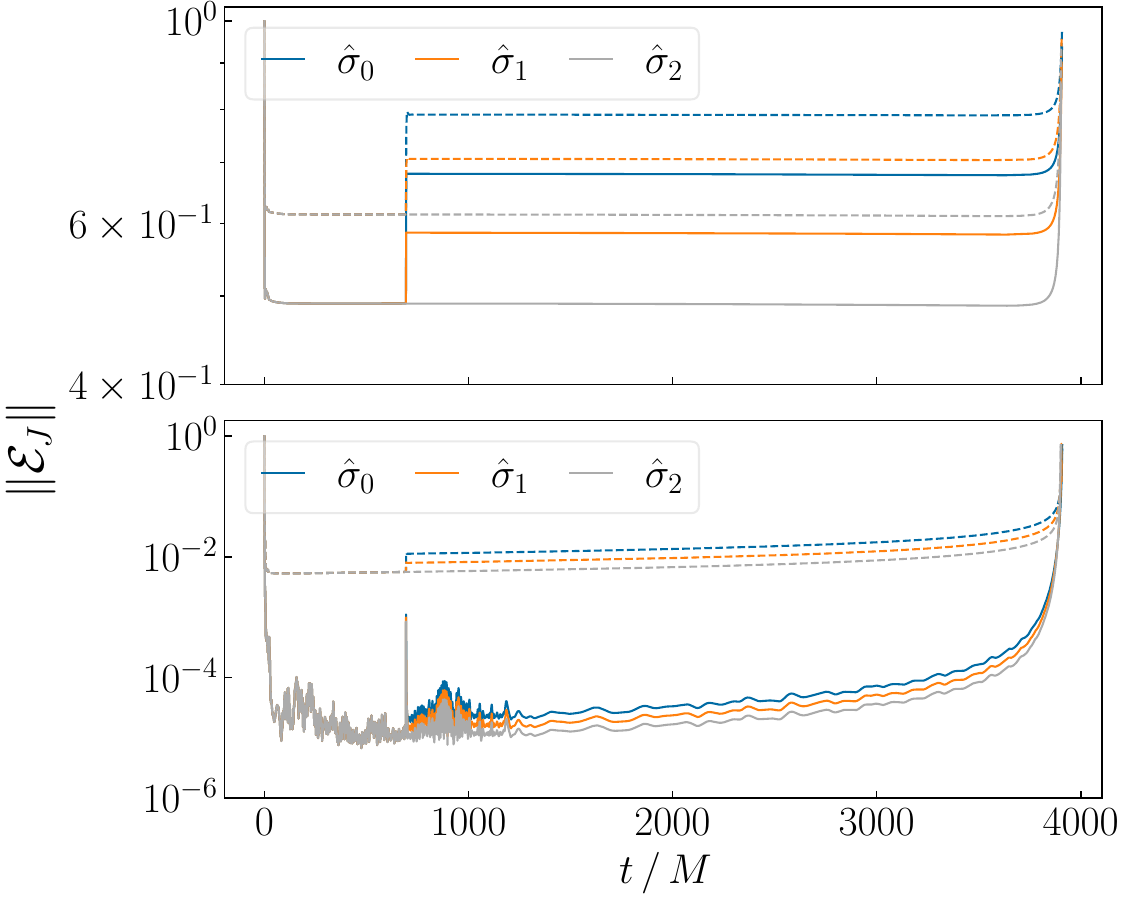}
    \caption{\emph{Tracking diagnostic with an advection driver.}
    \(L_2\)-norm of the scalar tracking diagnostic \(\mathcal{E}[\Sigma]\) (solid) and the tensor tracking diagnostic \(\mathcal{E}[\Sigma_{ab}]\) (dashed) over the entire spatial domain as we keep the resolution fixed to \(\mathrm{Lev}1\) but vary \(\hat{\sigma}\).
    Top: Using the advection driver [Eq.~\eqref{eq : advection driver}].
    Bottom:: Using the comoving driver CD1 [Eq.~\eqref{eq : comoving driver simplified}].
    Here, we have defined \(\{\hat{\sigma}_{-1} \equiv 4, \hat{\sigma}_0 \equiv 1, \hat{\sigma}_1 \equiv 1/4, \hat{\sigma}_2 \equiv 1/16\}\).
    The tracking diagnostics for the case of the advection driver are \(\sim 4\) orders of magnitude worse than the scalar tracking for the comoving driver case.
    }
    \label{fig : advection binary tracking}
\end{figure}
\begin{figure}
    \includegraphics[width=\linewidth]{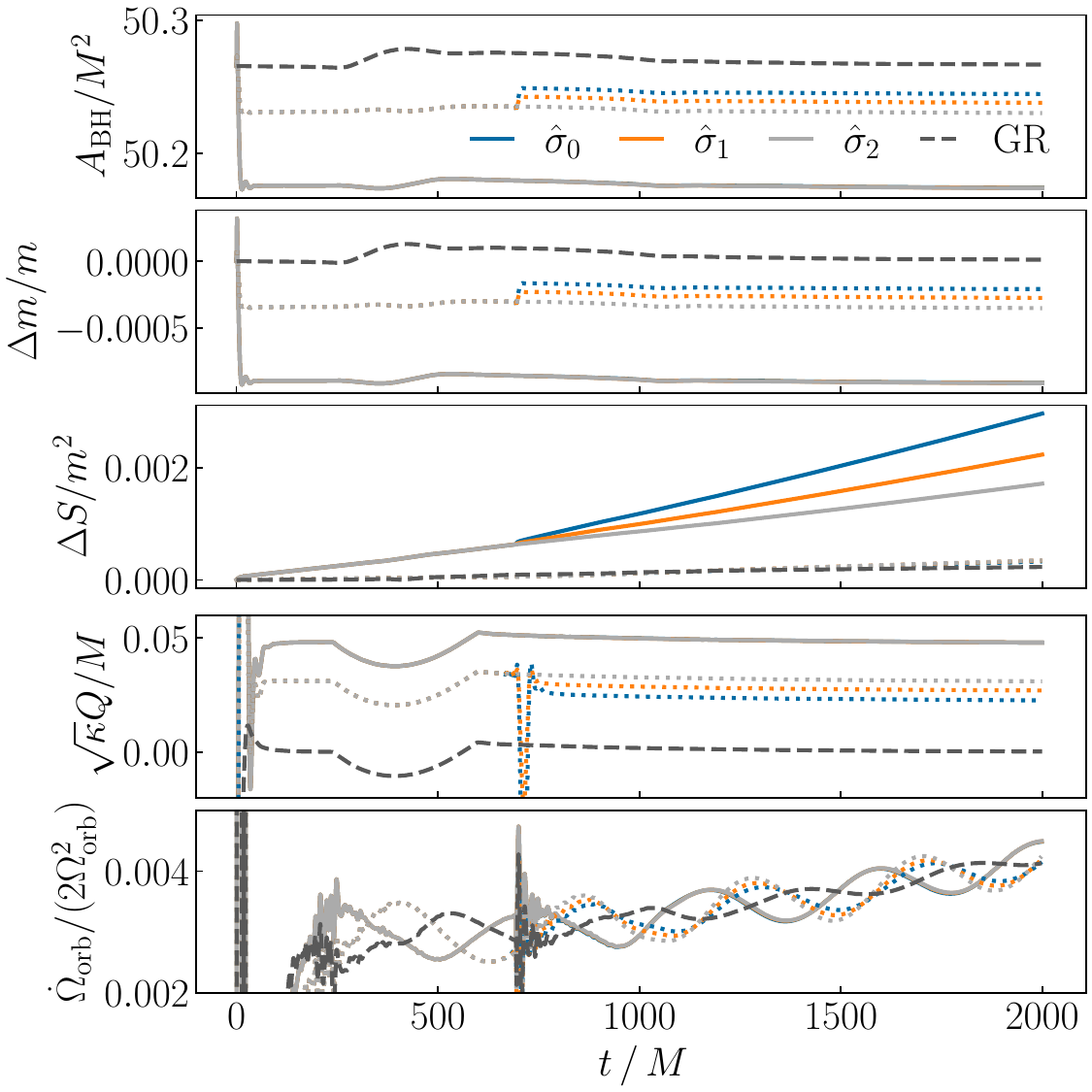}
    \caption{\emph{Time dependence of black hole and orbital quantities with an advection driver.}
    Shorter version (lower initial separation) of the simulations of Fig.~\ref{fig : binary bh area mass spin in time} compared against equivalent simulations using an advection driver for different values of \(\hat{\sigma}\).
    Here, \({\hat{\sigma}_{0} = 1, \hat{\sigma}_{1} = 1/4, \hat{\sigma}_{2} = 1/16}\).
    The advection driver [dotted; Eq.~\ref{eq : advection driver}] is not as effective as the comoving driver CD1 [solid; Eq.~\eqref{eq : comoving driver simplified}] at describing the area (first), mass (second) and scalar charge (fourth panel).
    Eccentricity is effectively reduced in both cases, as seen from the amplitude of the time derivative of the orbital frequency (bottom panel).
    As in the main text, the spin growth is addressed by switching to using the generalized driver CD2 [Eq.~\eqref{eq : comoving driver simplified second version}].
    }
    \label{fig : advection driver bh area mass spin in time}
\end{figure}

\section{Comparison with an advection driver \label{app : appendix comparison with and advection driver}}

In this appendix, we further examine the improvements of comoving drivers [Eq.~\eqref{eq : comoving driver simplified}] by directly comparing with simulations performed using the advection driver of Ref.~\cite{Cayuso:2023xbc}, and given by
\begin{align} \label{eq : advection driver}
    \sigma \left(\partial_t - \beta^{i}\partial_{i}\right)^{2} \boldsymbol{\Sigma} + \tau \left(\partial_t - \beta^{i}\partial_{i}\right) \boldsymbol{\Sigma} = - \left(\boldsymbol{\Sigma} - \boldsymbol{\mathcal{S}}\right).
\end{align}
Here, \(\beta^{i}\) is the shift vector in the \(3+1\)-decomposition ---see Eq.~\eqref{eq : spacetime metric 3plus1 decomposition}.
Eq.~\eqref{eq : advection driver} is a simplified version of Eq.~(8) in Ref.~\cite{Cayuso:2023xbc} that differs by factors of the lapse \(\alpha\) and non-principal part terms; these specific differences do not affect the conclusions below, with similar conclusions applying also for the case of wavelike drivers of the form~\eqref{eq : wavelike driver}.

In Fig.~\ref{fig : advection binary tracking}, we show the tracking diagnostics for the binary case of Sec.~\ref{sec : Results binary bbh CD1}.
For the purposes of this comparison, it suffices to start the binary evolution from a smaller initial separation and to compare only to the scalar comoving driver [Eq.~\eqref{eq : comoving driver simplified}].
The tracking diagnostics for the case of Eq.~\eqref{eq : advection driver} (top panel) are \(\sim 4\) orders of magnitude larger than the tracking diagnostic for the scalar variable (bottom panel) in the comoving driver case.
These large differences are also evident in the BH quantities early in the simulation (Fig.~\ref{fig : advection driver bh area mass spin in time}).
In particular, the overall amplitude of the area, mass, and scalar charge (which are mostly sensitive to the scalar part of driver equation) is particularly affected in the advection driver case (dotted lines).
Moreover, the dependence on the driver parameters \(\hat{\sigma}\) is more pronounced, since the advection driver does not have a stationary limit common to all values of \(\hat{\sigma}\) ---the difference arising from factor proportional to \(\sigma \partial^2_i \boldsymbol{\Sigma}\) and \(\sqrt{\sigma} \partial_i \boldsymbol{\Sigma}\).

To make sure this is not a normalization problem, we have checked that normalizing the charge by the Christodolou mass after the initial transient (so to as to account for the mass discrepancy), does not significantly reduce the discrepancy in the charge between the two drivers.

For reference, we have also compared these drivers for the case of single black holes in the test-field limit in App.~A of \paperone, including wavelike drivers of the form~\eqref{eq : wavelike driver}.


\section{No secular growth in amplitude \label{app : waveform length dependence}}

\begin{figure}
    \includegraphics[width=\linewidth]{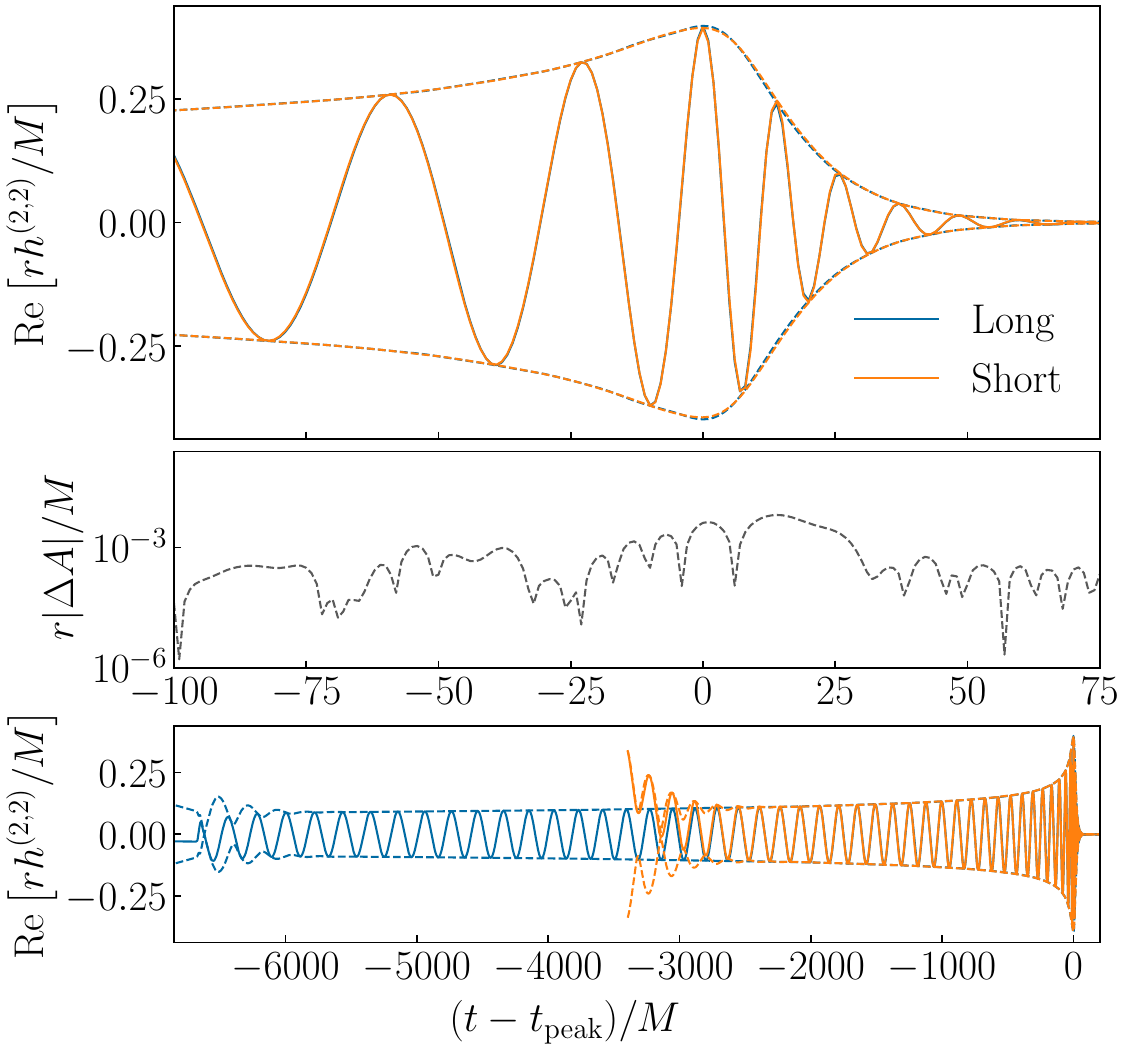}
    \caption{\emph{Dependence on the length of the waveform.}
    Top: The merger portion of the real part of the \(h^{(2,2)}\) mode is compared between a long and short simulations, with \(\sqrt{\kappa} \ell^2 / m^2  = 1/20\), \(\hat{\sigma} = \hat{\sigma}_2 = 1/16\) and \(\mathrm{Lev}1\) resolution.
    The waveforms are aligned in time and phase at the peak. We indicate the amplitude \(A\) with and envelope in dashed lines. 
    Middle: Difference in amplitude \(\Delta A\) between the two waveforms for the merger part. We see indication of secular growth in \(A\) between the two cases.
    Bottom: We show the entire wave form in both cases, for which CCE junk is visible at the start of the waveforms.
    }
    \label{fig : no secular growth}
\end{figure}

In this appendix, we investigate the effect of the length of the simulation on the gravitational waveform.
In Fig.~\ref{fig : no secular growth} we compare \(h^{(2,2)}\) for the simulations of Sec.~\ref{sec : Results binary bbh CD1} against the one obtained from simulations that are \(\sim 14\) orbits shorter.
The driver parameter is fixed to \(\hat{\sigma}  = \hat{\sigma}_2 = 1/16\), and \(\mathrm{L}1\), and the resolution to L1.
In this case, since we are only interested in the amplitude of the GW, we align the waveforms at the peak of the \((2,2)\)-mode and apply constant phase factor to make \(h^{(2,2)}\) real at the alignment time.
We find no evidence that secular growth affects our results:
we observe no apparent change in the amplitude, caused by accumulated  error from the additional 14 orbital cycles.
We contrast this to the order-reduction approach in previous attempts within the SXS Collaboration~\cite{Okounkova:2020rqw}, where secular errors grew quadraticaly with length, and where noticeable after \(\mathrm{O}(100 M)\).


\section{First-order system \label{app : details first order system}}

We recast Eqs.~\eqref{eq : fixed equations of motion} and Eq.~\eqref{eq : comoving driver simplified} as a first order system of the form
\begin{align}
    \partial_t \boldsymbol{u} + \mathbb{A}^{i} \left(\boldsymbol{u}\right) \partial_i \boldsymbol{u} = \boldsymbol{S}\left(\boldsymbol{u}\right).
\end{align}
where 
\begin{align}
    \boldsymbol{u} = \left(g_{ab}, \Pi_{ab}, \Phi_{iab}, \Psi, \Pi, \Phi_{i},\Sigma, \Pi^{(\Sigma)}, \Sigma_{ab}, \Pi^{(\Sigma)}_{ab}\right)
\end{align}
is a collection of first-order variables for the metric, scalar and auxiliary fields ---in total 77 evolved field components.
Here, we have defined
\begin{align} \label{eq : first order variables}
    \Pi_{ab} &\equiv -n^{c} \partial_{c}  g_{ab}, & \Phi_{iab} &\equiv \partial_i g_{ab},  \nonumber \\
    \Pi &\equiv -n^{c} \partial_{c} \Psi, & \Phi_{i} &\equiv \partial_{i} \Psi,  \nonumber \\
    \Pi^{(\Sigma)} & \equiv \mathcal{D}^{(s)}_t \Sigma, & \Pi^{(\Sigma)}_{ab} & \equiv \mathcal{D}^{(s)}_t \Sigma_{ab},
\end{align}
where \(n_a\) is the normal to the slicing,
and \(\mathcal{D}^{s}\) is either \((\partial_t + v^{i}\partial_{i})\) for the CD1 driver [Eq.~\eqref{eq : comoving driver simplified}], or \((\partial_t + \mathcal{L}_{v})\) for the CD2 driver [Eq.~\eqref{eq : comoving driver simplified second version}].
The first-order reduction introduces additional constraints
\begin{align} \label{eq : first order reduction constraints}
    \mathcal{C}^{}_{iab} \equiv \partial_{i} g_{ab} - \Phi_{iab} , &&
    \mathcal{C}^{(\Psi)}_{i} \equiv \partial_{i} \Psi - \Phi_{i} , \nonumber \\
    \mathcal{C}^{}_{ijab} \equiv 2 \partial_{[i} \Phi_{j]ab} , &&
    \mathcal{C}^{(\Psi)}_{ij} \equiv 2\partial_{[i} \Phi_{j]},
\end{align}
that we monitor during the evolution.

The first-order evolution system for the driver variables is
\begin{align}
    \mathcal{D}^{(s)}_{t} \Sigma &= - \alpha \Pi^{(\Sigma)} , \nonumber \\
    \sigma \mathcal{D}^{(s)}_{t} \Pi^{(\Sigma)}  &= -\alpha^{2} \tau \Pi^{(\Sigma)} + \alpha  \left(\Sigma - \mathcal{S}\right), \nonumber \\
    \mathcal{D}^{(s)}_{t} \Sigma_{ab} &= - \alpha \Pi^{(\Sigma)}_{ab} , \nonumber \\
    \sigma \mathcal{D}^{(s)}_{t} \Pi^{(\Sigma)}_{ab} &= -\alpha^{2} \tau \Pi^{(\Sigma)}_{ab} + \alpha \left( \Sigma_{ab} - \mathcal{S}_{ab}\right) .
\end{align}
The source terms \(\mathcal{S}\) in terms of as computed from \((3+1)\) and first-order variabels is presented in Appendix~\ref{app : source term expressions}.

\section{Expressions for the source terms \label{app : source term expressions}}

In this appendix, we provide explicit expressions for the source terms \(\boldsymbol{\mathcal{S}}\) [appearing in Eqs.~\eqref{eq : equations of motion} and Eqs.~\eqref{eq : fixed equations of motion}] as they were implemented in our code.
We have found it useful to express these quantities in terms of the electric and magnetic parts of the Weyl tensor where possible.


\subsection{\((3+1)\)-decomposition}

We decompose the spacetime metric in the \((3+1)\)-decomposition as
\begin{align} \label{eq : spacetime metric 3plus1 decomposition}
    ds^2 = -\alpha^{2} \, dt^2  + \gamma_{ij} \left(\beta^{i} \, dt + dx^{i}\right) \left(\beta^{j} \, dt + dx^{j}\right) .
\end{align}
where \(\alpha\) is the lapse, \(\beta^{i}\) is the shift, and \(\gamma_{ij}\) is the spatial metric.
The normal to the foliation is \(n_a = (-\alpha, \boldsymbol{0})\).
In practice, we compute \(\{\alpha, \beta^{i}, \gamma_{ij}\}\) from the GH variables [Eq.~\eqref{eq : first order variables}].


\subsection{Curvature quantities}

In the following, whenever there could be ambiguity, we refer to the curvature quantities computed in terms of the spatial or spacetime metrics with superscipts \(^{(3)}\) or \(^{(4)}\), respectively.
For example, the spatial Christoffel symbols are
\begin{align}
    {^{(3)} \Gamma^{k}_{ij}} = \frac{1}{2} \gamma^{kl} \left(\partial_{i} \gamma_{jl} + \partial_{j} \gamma_{il} - \partial_{l} \gamma_{ij}\right),
\end{align}
whereas the spacetime Christoffel symbols are
\begin{align}
    {^{(4)} \Gamma^{c}_{ab}} = \frac{1}{2} g^{cd} \left(\partial_{a} g_{bd} + \partial_{b} g_{ad} - \partial_{d} g_{ab}\right).
\end{align}

The extrinsic curvature is defined as
\begin{align}
    K_{ab} \equiv - \dfrac{1}{2} \mathcal{L}_{n} \gamma_{ab}.
\end{align}
It is a spatial tensor and is related to the first-order GH variables [Eq.~\eqref{eq : first order variables}] by
\begin{align}
    K_{ij} =  \dfrac{1}{2} \Pi_{ij} + \Phi_{(ij)a}n^a.
\end{align}
Spatial partial derivatives of \(\{\alpha, \beta^{i}, \gamma_{ij}\}\) are computed from similar expressions in terms of GH variables.

We compute the spatial Ricci tensor from
\begin{multline}
    {^{(3)} R}_{ab} = \partial_c {^{(3)}\Gamma}^{c}_{ab} - \partial_{(b} {^{(3)}\Gamma}^{c}_{a)c} \\
    + {^{(3)}\Gamma}^{d}_{ab}{^{(3)}\Gamma}^{c}_{cd} - {^{(3)}\Gamma}^{d}_{ac} {^{(3)}\Gamma}^{c}_{bd} .
\end{multline}

Lastly, we numerically evaluate \(\partial_{i}K_{jk}\) and \(\partial_{i}{^{(3)} \Gamma}_{jk}\) with spectral methods.
While we could express more of these expressions in terms of GH variables, at this stage, we have not optimized these calculations.


\subsection{Electric and Magnetic Parts}

The electric \(E_{ab}\) and magnetic \(B_{ab}\) parts of the Weyl tensor \(C_{abcd}\) are
\begin{align}
    E_{ab} &\equiv n^{c} n^{d} {C}_{acbd},\\
    B_{ab} &\equiv - n^{c} n^{d} {^{*} C}_{acbd} ,
\end{align}
where \({^{*}C}_{abcd} \equiv \frac{1}{2} \epsilon_{abef} {C^{ef}}_{cd}\) is the left-dual tensor.
Both \(E_{ab}\) and \(B_{ab}\) are symmetric spatial tensors.

The electric part of the Weyl tensor is
\begin{multline}
    E_{ab} = {^{(3)}R}_{ab} + \gamma^{cd}  (K_{cd} K_{ab} - K_{ad} K_{cb}) - \dfrac{1}{2} \gamma^{c}_{a} \gamma^{d}_{b} {^{(4)}R}_{cd} \\
    - \dfrac{1}{2} {^{(4)}R}_{cd} \gamma^{cd} \gamma_{ab} + \dfrac{1}{3} {^{(4)}R} \gamma_{ab}.
\end{multline}
Using Eq.~\eqref{eq : fixed equations of motion},
\begin{multline} \label{eq : weyl electric sigma}
    E_{ab} = {^{(3)}R}_{ab} + \gamma^{cd}  (K_{cd} K_{ab} - K_{ad} K_{cb}) - \dfrac{1}{2} \gamma^{c}_{a} \gamma^{d}_{b} \Sigma_{cd} \\
    - \dfrac{1}{2} \Sigma_{cd} \gamma^{cd} \gamma_{ab} + \dfrac{1}{3} \Sigma \gamma_{ab}.
\end{multline}
The electric part is traceless if it satisfies the Hamiltonian constraint~\cite{Alcubierre:1138167}.
In our evaluation scheme, we explicitly enforce tracelessness by
\begin{align}
    E_{ij} \to E_{ij} - \dfrac{1}{3}\gamma^{kl}E_{kl}\gamma_{ij}.
\end{align}

The magnetic part is given by
\begin{align} \label{eq : weyl magnetic sigma}
    B_{ij} = \dfrac{1}{\sqrt{\gamma}} \, D_{k}K_{l(i}\gamma_{j)m} \varepsilon^{mlk} ,
\end{align}
where \(D\) is the spatial covariant derivative, \(\gamma \equiv \det(\gamma_{ij})\), and \(\varepsilon^{ijk}\) is the Levi-Civita \emph{symbol} in 3 dimensions.
The magnetic part is already symmetric and traceless.


\subsection{The interaction tensor \(\ell^2 H^{\mathrm{(TR)}}_{ab}\)}

We find it useful to split the computation of \(H_{ab}\) [defined in Eq.
\eqref{eq : THP definitions}] in two parts
\begin{align}
    H^{}_{ab} = -8 (\hat{H}^{(C)}_{ab} + \hat{H}^{(R)}_{ab}),
\end{align}
where
\begin{align} \label{eq : weyl and ricci parts of H}
    \hat{H}^{(C)}_{ab} &\equiv C_{acbd}\nabla^{c}\nabla^{d}f(\Psi) \\
    \hat{H}^{(R)}_{ab} &\equiv (P_{acbd} - C_{acbd})\nabla^{c}\nabla^{d}f(\Psi) ,
\end{align}
and \(P_{acbd}\) was defined in Eq.~\eqref{eq : THP definitions}.

To evaluate the tensor source terms \(\mathcal{S}_{ab}\), we then compute the trace-reverse \(H^{\mathrm{(TR)}}_{ab} \equiv H_{ab} - (1/2)g^{cd}H_{cd} g_{ab} \).
Both, however, require us to compute first the double covariant derivative of the sGB shape function \(\nabla_{a} \nabla_{b} f(\Psi)\).


\subsubsection{The \(\nabla \nabla f\) tensor}

The double covariant derivative \(\nabla_{a} \nabla_{b} f(\Psi)\) can be related to the covariant double derivative of \(\Psi\) by the chain rule,
\begin{align}
    \nabla_{a} \nabla_{b} f (\Psi)&= f''(\Psi) \partial_{a} \Psi \partial_{b} \Psi
    + f'(\Psi) \nabla_{a} \nabla_{b} \Psi
    .
\end{align}

In order to evaluate \(\nabla_{a} \nabla_{b} f(\Psi)\), we need to compute the spacetime derivative of the scalar \(\partial_{a} \Psi\) and \(\nabla_{a} \nabla_{b} \Psi\). 
The former can be written in terms of our first-order scalar variables as
\begin{align}
    \partial_{a} \Psi = \gamma^{b}_{a}\Phi_{b} + n_{a} \Pi ,
\end{align}
with components \((\partial_{0} \Psi, \partial_{i} \Psi) = (- \alpha \Pi + \beta^{j} \Phi_{j}, \Phi_{i})\).

We construct \(\nabla_{a} \nabla_{b} \Psi\) from its projections (App.~\ref{sec: four tensors from projections}) (see Sec.~\ref{sec: four tensors from projections}),
\begin{align}
    \rho^{(\nabla \nabla \Psi)} &\equiv n^{a} n^{b} \nabla_{a} \nabla_{b} \Psi \nonumber \\
    &= - \mathcal{L}_{n} \Pi - \dfrac{1}{\alpha} \gamma^{ij}\Phi_{i} \partial_{j} \alpha
    , \nonumber \\
          &= -\dfrac{1}{\alpha}(\partial_t \Pi - \beta^{i} \partial_{i} \Pi + \gamma^{ij} \Phi_{i} \partial_{j} \alpha ) ,
    \\
    {j^{(\nabla \nabla \Psi)}}_{i} &\equiv \gamma_{i}^{a} n^{b} \nabla_{a} \nabla_{b} \Psi \nonumber \\
    &= {K_{i}}^{j} \Phi_{j} - D_i \Pi
    , \nonumber \\
    {S^{(\nabla \nabla \Psi)}}_{ij} &\equiv \gamma_{i}^{a} \gamma_{j}^{b} \nabla_{a} \nabla_{b} \Psi \nonumber \\
    &= D_{i} \Phi_{j} - \Pi K_{ij}    \\
    &= -\Pi K_{ij}
          + \dfrac{1}{2} (\partial \Phi_{ij} + \partial \Phi_{ji}) - {^{(3)}\Gamma}^{k}_{ij} \Phi_{k},
\end{align}
where \(\partial_t \Pi\) is computed using the scalar equation of motion (and sourced by \(\Sigma\)).


\subsubsection{Weyl part}

Here, we describe how to compute \(\hat{H}^{(C)}_{ab}\) in Eq.~\eqref{eq : weyl and ricci parts of H}. 
We first decompose the Weyl tensor in terms of \(\{E_{ab}, B_{ab}\}\) as~\cite{Alcubierre:1138167}
\begin{align}
    C_{abcd} &=   E_{ac} \left( \gamma_{bd} + n_{b} n_{d} \right) { - \epsilon_{abe} n_{d} {B_{c}}^{e} }
    \nonumber \\
    &\quad - E_{bc} \left( \gamma_{ad} + n_{a} n_{d} \right) { - \epsilon_{cde} n_{b} {B_{a}}^{e} }
    \nonumber\\
    &\quad  - E_{ad} \left( \gamma_{bc} + n_{b} n_{c} \right) { + \epsilon_{abe} n_{c} {B_{d}}^{e}}
    \nonumber\\
    &\quad  + E_{bd} \left( \gamma_{ac} + n_{a} n_{c} \right) { + \epsilon_{cde} n_{a} {B_{b}}^{e} } .
\end{align}
we then construct the Weyl part from its (3+1)-projections (Sec.~\ref{sec: four tensors from projections})
\begin{align}
    \rho^{(H)} &\equiv n^{a} n^{b} \hat{H}^{(C)}_{ab} \nonumber \\
    &= E_{ij} {S^{(\nabla \nabla f)}}^{ij} ,\\
    {j^{(H)}}_{i} &\equiv n^{a} \gamma^{b}_{i} \hat{H}^{(C)}_{ab} \nonumber  \\
    &= E_{ij} {j^{(\nabla \nabla f)}}^{j}  { -  \sqrt{\gamma} \, \mathrm{tr}_{\boldsymbol{\gamma}}(\boldsymbol{S} \times \boldsymbol{B})_{i}} ,
    \\
    {S^{(H)}}_{ij} &\equiv \gamma^{a}_{i} \gamma^{b}_{j} \hat{H}^{(C)}_{ab} \nonumber  \\
    &= \left({S^{(\nabla \nabla f)}}_{kl} \gamma^{lk} + \rho^{(\nabla \nabla f)}\right) E_{ij} 
    \nonumber \\
    & \quad
    - (
    {S^{(\nabla \nabla f)}}_{jk} \gamma^{kl} E_{li} 
    + {S^{(\nabla \nabla f)}}_{ik} \gamma^{kl} E_{lj})
    \nonumber \\
    & \quad 
     {- \sqrt{\gamma} \left[
          (\boldsymbol{j} \times \boldsymbol{B})_{ij} + (\boldsymbol{j} \times \boldsymbol{B})_{ji}
          \right] }
    + \gamma_{ij} \rho^{(H)} ,
\end{align}
where
\begin{align}
    \mathrm{tr}_{\boldsymbol{\gamma}}(\boldsymbol{S} \times \boldsymbol{B})_{i} &\equiv                 \varepsilon_{ijk}  {S^{(\nabla \nabla f)}}^{jl}
     B_{l}^{k} , \nonumber \\
    (\boldsymbol{j} \times \boldsymbol{B})_{il} &\equiv
    \varepsilon_{ijk}  {j^{(\nabla \nabla f)}}^{j} B_{l}^{k} .
\end{align}


\subsubsection{Ricci part}

Here, we describe how to compute \(\hat{H}^{(R)}_{ab}\) in Eq.~\eqref{eq : weyl and ricci parts of H}. 

Decomposing the Riemann tensor in terms of the Weyl and the 4-Ricci tensors we obtain
\begin{multline}
    \hat{H}^{(R)}_{ab} = \Big\{
        \dfrac{1}{2} \Big(-g_{cd} {^{(4)}R}_{ab} + g_{bc} {^{(4)}R}_{ad} \\
    + g_{ad} {^{(4)}R}_{bc} - g_{ab}{^{(4)}R}_{cd}\Big) \\
    + \dfrac{1}{3} (-g_{ad} g_{bc} + g_{ab} g_{cd}) {^{(4)}R}
    \Big\} \nabla^{c} \nabla^{d} f (\Psi)
\end{multline}
Using Eq.~\eqref{eq : fixed equations of motion},
\begin{multline}
    \hat{H}^{(R)}_{ab} = \Big\{
    \dfrac{1}{2} \Big(-g_{cd} \Sigma_{ab} + g_{bc} \Sigma_{ad} \\
    + g_{ad} \Sigma_{bc} - g_{ab}\Sigma_{cd}\Big) \\
    + \dfrac{1}{3} (-g_{ad} g_{bc} + g_{ab} g_{cd}) g^{ef}\Sigma_{ef}
    \Big\} \nabla^{c} \nabla^{d} f (\Psi) .
\end{multline}


\subsection{Four-tensors from projections \label{sec: four tensors from projections}}

We can reconstruct a 4-tensor from its projections as
\begin{align}
    T_{ab} = \rho^{(T)} n_{a}n_{b} - 2 \gamma^{i}_{(a}  n_{b)} j^{(T)}_{i} + \gamma^{i}_{a} \gamma^{j}_{b} S^{(T)}_{ij}.
\end{align}

In terms of 4-spacetime components
\begin{align}
    T_{00} &= \alpha^2 \rho^{(T)} + 2 \alpha \beta^{i} j^{(T)}_{i} + \beta^{i} \beta^{j} S^{(T)}_{ij}, \nonumber\\
    T_{0k} &= \alpha j^{(T)}_{k} + \beta^{i}S^{(T)}_{ik}, \nonumber \\
    T_{ij} &= S^{(T)}_{ij}.
\end{align}
Here, we have defined the projections \(\rho^{(T)} \equiv n^{a}n^{b}T_{ab}\), \(j^{(T)}_{i} \equiv n^{a}\gamma^{b}_{i} T_{ab}\) and \(S^{(T)}_{ij} \equiv \gamma^{a}_{i}\gamma^{b}_{j}T_{ab}\).

\subsection{Scalar source \label{sec: appendix scalar source}}

Here, we describe how to compute \(\mathcal{S}\) in Eq.~\eqref{eq : source terms abstract}. 
We decompose the Riemann tensor in terms of the Weyl and the 4-Ricci tensors we can write the Gauss-Bonnet scalar can be rewritten in terms \(\{E_{ab}, B_{ab}\}\) as
\begin{align}
    \mathcal{G} &=
    R_{abcd}R^{abcd} - 4 {^{(4)}}R_{ab} {^{(4)}}R^{ab} + {^{(4)}}R^2  \nonumber\\
    &= 8 \left(E_{ab}E^{ab} - B_{ab} B^{ab} \right) 
        -2 {^{(4)}}R_{ab}{^{(4)}}R^{ab} + \dfrac{2}{3} {^{(4)}}R^2 .
\end{align}
Using Eq.~\eqref{eq : fixed equations of motion},
\begin{align}
    \mathcal{G} 
    &= 8 \left(E_{ab}E^{ab} - B_{ab} B^{ab} \right) 
        -2 \Sigma_{ab}\Sigma^{ab} + \dfrac{2}{3} (g^{ab}\Sigma_{ab})^2  .
\end{align}
We then use the expressions in Eq.~\eqref{eq : weyl electric sigma} and \eqref{eq : weyl magnetic sigma} for the electric and magnetic parts.


\FloatBarrier


\bibliography{paper}

\end{document}